\documentclass[final,5p,times,twocolumn]{elsarticle}

\pdfoutput=1

\biboptions{numbers,sort&compress}

\usepackage{amsmath}
\usepackage{amsthm}
\newtheorem{definition}{Definition}
\usepackage{algorithmic}

\usepackage{url}

\usepackage[caption=false,font=footnotesize]{subfig}

\usepackage{tikz}
\usetikzlibrary{arrows,decorations.pathreplacing,shapes.misc,patterns,shapes,backgrounds,matrix,positioning,chains}
\tikzset{cross/.style={cross out, draw=black, minimum size=2*(#1-\pgflinewidth), inner sep=0pt, outer sep=0pt},
	cross/.default={1pt}}

\def\firstcircle{(0,0) circle (1.2cm)}
\def\secondcircle{(0.4,0.7) circle (1.2cm)}
\def\thirdcircle{(0.3,1.4) circle (1.2cm)}
\def\fourthcircle{(1.25,0.45) circle (1.2cm)}

\usepackage{soul}

\usepackage{multirow}
\usepackage{hhline}

\newtheoremstyle{tcdefinition}
{2pt}
{2pt}
{\itshape}
{0pt}
{\bfseries}
{:}
{ }
{\thmname{#1}\thmnumber{ #2 }\thmnote{\mdseries{[#3]}}}
\theoremstyle{tcdefinition}
\newtheorem{tcdefinition}{Termination Criterion}[]

\usepackage{color}
\definecolor{forestgreen}{rgb}{0.13, 0.54, 0.13}
\definecolor{beaublue}{rgb}{0.74, 0.83, 0.9}
\definecolor{amaranth}{rgb}{0.9, 0.17, 0.31}
\definecolor{airblue}{rgb}{0.36, 0.54, 0.66}
\definecolor{lavendergray}{rgb}{0.77, 0.76, 0.82}
\definecolor{isabelline}{rgb}{0.96, 0.94, 0.93}
\definecolor{hansayellow}{rgb}{0.91, 0.84, 0.42}
\definecolor{indianyellow}{rgb}{0.89, 0.66, 0.34}
\definecolor{mikadoyellow}{rgb}{1.0, 0.77, 0.05}
\definecolor{amber}{rgb}{1.0, 0.49, 0.0}
\definecolor{amethyst}{rgb}{0.6, 0.4, 0.8}

\journal{Computer Communications}

\begin{document}

\begin{frontmatter}



\title{Revisiting XOR-based Network Coding for Energy Efficient Broadcasting in Mobile \\Ad Hoc Networks}



\author[myu]{Nikolaos~Papanikos}
\ead{npapanik@cse.uoi.gr}

\author[myu]{Evangelos~Papapetrou\corref{cor1}}
\ead{epap@cse.uoi.gr}

\address[myu]{Department of Computer Science and Engineering, University of Ioannina, GR 45110, Greece}

\cortext[cor1]{Corresponding author}

\begin{abstract}
Network coding is commonly used to improve the energy efficiency of network-wide broadcasting in wireless multi-hop networks. In this work, we focus on XOR-based broadcasting in mobile ad hoc networks with multiple sources. We make the observation that the common approach, which is to benefit from the synergy of XOR network coding with a CDS-based broadcast algorithm, suffers performance breakdowns. After delving into the details of this synergy, we attribute this behavior to an important mechanism of the underlying broadcast algorithm, known as the ``termination criterion''. To tackle the problem, we propose a termination criterion that is fully compatible with XOR coding. In addition to that, we revisit the internals of XOR coding. We first enhance the synergy of XOR coding with the underlying broadcast algorithm by allowing each mechanism to benefit from information available by the other. In this way, we manage to improve the pruning efficiency of the CDS-based algorithm while at the same time we come up with a method for detecting coding opportunities that has minimal storage and processing requirements compared to current approaches. Then, for the first time, we use XOR coding as a mechanism not only for enhancing energy efficiency but also for reducing the end-to-end-delay. We validate the effectiveness of our proposed algorithm through extensive simulations on a diverse set of scenarios.  
\end{abstract}


\end{frontmatter}

\section{Introduction}\label{sec:introduction}
Network-layer broadcasting is fundamental for mobile ad hoc networks because it provides the means to disseminate information throughout the network. Besides application data, broadcast protocols are also used to distribute control information to every network node~\cite{abolhasan-review,Mian2009-servicedisc}. In this way, each node maintains a view of the network structure that is a basic element for many networking mechanisms. The seminal work by Ahlswede et al.~\cite{Ahlswede} introduced network coding, a concept that significantly enhances the performance of networking protocols in both wired and wireless networks. As a result, over the last years, many researchers focused on incorporating network coding into broadcasting in wireless ad hoc networks~\cite{CodeB,NCDS_conf,NCDS_journal,kunz-iwcmc,OstovariXOR,Directional,fragouli_rlnc,widmer-extreme-net,mahmood_arlnccf_icc,RLDP,DiSC,WangXOR,rahnavard2008CRBCast,vellambi2010FTS,widmer_rlnc-update,hou2008adapcode,cho_DRAGON,yang2011-RCODE-jrn,subramanian2012uflood,Chachulski-more,Koutsonikolas-pacifier,OstovariRLNC}. Based on their main objective, coding-based approaches can be classified into: i) energy-efficient, and ii) delivery guarantee schemes. The schemes of the first class~\cite{CodeB,NCDS_conf,NCDS_journal,kunz-iwcmc,OstovariXOR,Directional,fragouli_rlnc,widmer-extreme-net,mahmood_arlnccf_icc,RLDP} utilize network coding towards energy efficiency aiming to strike the best possible balance between delivery and cost (as expressed by the number of transmissions). On the other hand, the schemes of the second class~\cite{DiSC,WangXOR,rahnavard2008CRBCast,vellambi2010FTS,widmer_rlnc-update,hou2008adapcode,cho_DRAGON,yang2011-RCODE-jrn,subramanian2012uflood,Chachulski-more,Koutsonikolas-pacifier,OstovariRLNC} use network coding to guarantee the delivery of broadcast packets to all network nodes, treating the minimization of the related costs as a secondary objective.

In this work we focus on energy-efficient broadcasting in mobile ad hoc networks. Moreover, we are interested in the scenario of multiple broadcasting sources, i.e. we examine the many-to-all and all-to-all communication paradigms. Such scenarios appear rather frequently when multiple nodes in parallel and independently engage in discovery phases. Some representative examples include discovering routes in on-demand routing protocols~\cite{abolhasan-review}, locating resources in service discovery applications~\cite{Mian2009-servicedisc, Ververidis-servicedisc} and retrieving volatile data from peer databases ~\cite{Wolfson-peerdb,aggelidis-globecom}. In all of the aforementioned examples the focus is on energy efficiency rather than on guaranteeing delivery.

In the field of energy efficient broadcasting the most popular design choice is to adopt XOR-based network coding~\cite{COPE}. Algorithms that follow this approach~\cite{CodeB,NCDS_conf,NCDS_journal,kunz-iwcmc,OstovariXOR,Directional} encode packets on a hop-by-hop basis using bitwise XOR and then forward them using a Connected Dominating Set (CDS) based broadcasting scheme~\cite{CDS-study,MPR,DP,PDP}. Although this strategy has been proved successful, we bring to light several occasions where its performance severely degrades and the coding gain becomes negligible. Motivated by this finding, we examine in depth the synergy of network coding and the underlying CDS-based broadcast algorithm. We conclude that the weak link is the mechanism of the broadcasting algorithm known as ``the termination criterion''. Therefore, as a first step, we explore the use of alternative termination criteria proposed in the literature of traditional broadcasting. Unfortunately, we find that none of them is compatible with XOR network coding. To address the problem, we propose the \emph{Network Coding Broadcast with Coded Redundancy} (NOB-CR) algorithm which incorporates a novel termination criterion that is fully compatible with XOR coding. Moreover, NOB-CR revisits the coding internals in order to enhance the overall performance in terms of energy efficiency, delivery delay and utilization of network resources. In summary, our main contributions are:
\begin{itemize}
	\item We unveil the shortcomings in the synergy between XOR coding and CDS-based broadcasting (Section~\ref{sec:synergy}). Then, after analyzing the reasons of this finding, we propose a coding-friendly termination criterion for the CDS-based algorithm and illustrate its efficiency (Section~\ref{sec:MCU}).
	\item We delineate a novel method for detecting coding opportunities (Section~\ref{sec:coding_opp}). The method is lightweight in the sense that, in contrast to current approaches, requires each node to maintain minimum state while it mostly utilizes information that is already available through the underlying broadcast mechanism.
	\item We enhance the pruning efficiency of the underlying CDS-based algorithm by exploiting information available from the coding mechanism, i.e. we establish a bidirectional synergy between network coding and CDS-based forwarding (Section~\ref{sec:Pruning}).
	\item We address the problem of increased end-to-end delay in network coding broadcasting (Section~\ref{sec:coded_redundancy}). This problem is a direct consequence of Random Assessment Delay (RAD), a mechanism used by XOR coding in an effort to increase coding opportunities. The proposed solution, called \emph{Coded Redundancy}, takes a novel approach and uses XOR coding to achieve a cost-free increase of packet redundancy across the network in order to reduce end-to-end delay.
\end{itemize}

\noindent The rest of the paper is organized as follows. In Section~\ref{sec:Preliminaries}, we review the basic principles of CDS-based broadcasting and XOR network coding. In Section~\ref{sec:Evaluation}, we present an extensive evaluation of the proposed algorithm. In Section~\ref{sec:Related}, we review the existing work related to coding-based broadcasting in wireless ad hoc networks. Finally, we summarize our findings in Section~\ref{sec:Conclusion}.

\section{Preliminaries}\label{sec:Preliminaries}
Before continuing, we first briefly review the basic principles of CDS based broadcasting as well as XOR-based coding.   

\subsection{CDS-based Broadcast Principles}\label{sec:brodcasting_specs}
Energy efficient broadcast algorithms aim to minimize the number of transmissions required for delivering a packet to all network nodes~\cite{survey-Williams,survey-Ruiz}. The most effective algorithms follow the CDS-based broadcasting approach. According to this, the algorithm constructs a connected dominating set of the network~\cite{CDS-study,MPR,DP}. The nodes constituting the CDS are the \emph{forwarders}, i.e. those elected to forward the broadcast packets, while all other nodes just act as passive receivers. Since computing the forwarders should be performed in a distributed fashion, the common approach is to approximate them locally at each node $v$ using its 1-hop neighbor set ($\mathcal{N}(v)$), i.e. the set that consists of $v$'s one hop neighbors, and the 2-hop neighbor set ($\mathcal{N}(\mathcal{N}(v))$), i.e. the set consisting of all nodes that lie at maximum two hops away from $v$. 

Even though transmitting packets only through forwarders successfully reduces packet duplicates, a significant number of them still exists across the network. This is because the selection of forwarders is made in a distributed manner and with limited information. As a result, special attention should be given to these duplicates as they could lead to additional transmissions and degrade energy efficiency. Therefore, the reception of a packet duplicate in a forwarder node leads to a dilemma whether to forward it or not. Forwarding the duplicate could increase redundant transmissions while dropping it could potentially impact the delivery efficiency. The mechanism that is responsible to handle such situations is the \emph{termination criterion}. Multiple criteria have been proposed in the literature ~\cite{PDP,TC-CU,TC_MRU,MPR2}. In the rest of this paper we will use the terminology proposed in \cite{PDP} and \cite{TC-CU} to refer to these criteria:
\begin{tcdefinition}[Marked/unmarked (M/U)]
Each node keeps track of the packets received by each of its 1-hop neighbors. Then, in the case of a duplicate reception, a forwarder transmits the received duplicate if at least one of the neighbors is not marked to have received the packet.
\end{tcdefinition}	
\noindent This is the most well-known approach. However, having all nodes to store the reception status for all of their 1-hop neighbors and for all packets could be a daunting challenge in terms of both memory usage and processing overhead~\cite{PDP}.
\begin{tcdefinition}[Relayed/unrelayed (R/U)]
	A forwarder transmits a duplicate only if no other duplicate of the same packet has been relayed by the same forwarder in the past.
\end{tcdefinition}
\begin{tcdefinition}[Covered/uncovered (C/U)]
	A node acting as a forwarder relays packets seen for the first time while it drops already seen packets including the ones not relayed in the past because the node was not elected as a forwarder at that time.
\end{tcdefinition}
\noindent In contrast to M/U, the latter two approaches are more realistic due to the limited storage and processing requirements.

The algorithms proposed in the literature follow two major strategies for building the CDS, i.e. calculating the forwarders. 
The first is to build a CDS that is common to every network node using local information~\cite{WuLi,DaiWu,it-cds,MPR-survey,SBA,DS-NES,DS-NES-supplementary,PBSM,ABSM,MPR,MPR2,ABSM-theoretical,MPR-reliable} while the second is to build a source-specific CDS~\cite{DP,PDP,Khabbazian,DCB}. In the first category the nodes of the CDS are used for any packet regardless of its source and updated whenever topology changes are detected. Most efficient studies in this line of research also use information related to the broadcast process, e.g. packet reception status, in order to further prune transmissions and/or enhance reliability~\cite{SBA,DS-NES,DS-NES-supplementary,PBSM,ABSM,MPR,MPR2,ABSM-theoretical,MPR-reliable}.
On the other hand, in the second category, a node that relays a packet calculates the list of forwarders by considering the previous hop of the packet and piggybacks the corresponding list on it. In this way, a source-based CDS is formed for each packet. More specifically, when a node $v$ receives a packet from $u$ checks whether it is selected as a forwarder. If so, a common approach is to elect forwarders so as to deliver the packet to (or ``cover") the set $\mathcal{U}(v)$ of nodes that lie exactly 2-hops away from $v$, i.e. $\mathcal{U}(v)\!\!=\!\!\mathcal{N}(\mathcal{N}(v))\!-\!\mathcal{N}(v)$. The set of candidate forwarders $\mathcal{C}(v)$ is in general a subset of $v$'s neighbors, i.e. $\mathcal{C}(v)\!\!\subseteq\!\!\mathcal{N}(v)$. Note that $\mathcal{U}(v)\!\!\subseteq\!\bigcup_{\forall u\in \mathcal{C}(v)}\mathcal{N}(u)$ and that $\mathcal{C}(v)$ can be seen as a set of sets if each node $u\in \mathcal{C}(v)$ is replaced by $\mathcal{N}(u)$, thus the election of forwarders is modeled as a set cover problem. The solution is usually given by the well-known greedy set cover (GSC) algorithm~\cite{intro-algo-cormen}, however other more efficient approximation algorithms exist~\cite{Sarac-TPDS,Sarac-AdHocNets,MPR,epap-efcn}. Furthermore, node $v$ takes advantage of $u$'s neighborhood to reduce both the set of candidate forwarders, i.e. $\mathcal{C}(v)\!\!=\!\!\mathcal{N}(v)\!-\!\mathcal{N}(u)$, and the set of nodes $\mathcal{U}(v)$ that should receive the packet. Algorithms in the sourced-based CDS category vary in the approach taken to minimize the set $\mathcal{U}(v)$ and therefore the number of forwarders. TDP and PDP~\cite{PDP} exploit $u$'s two-hop neighborhood and further minimize the $\mathcal{U}(v)$ set. For example, node $v$ in PDP elects forwarders in order to cover the nodes in $\mathcal{U}(v)\!=\!\mathcal{N}(\mathcal{N}(v))\!-\!\mathcal{N}(v)\!-\!\mathcal{N}(u)\!-\!\mathcal{N}(\mathcal{N}(u)\cap \mathcal{N}(v))$. In the context of this paper, for presentation purposes we selected Partial Dominant Pruning (PDP)~\cite{PDP} as the reference algorithm. However, our findings can be easily generalized to all CDS-based broadcast protocols. The reason is that, as far as the CDS-based operation is concerned, we focus on the termination criterion which is a generic mechanism that does not depend on the specifics of each algorithm.

\subsection{XOR Coding Specifics}\label{sec:coding_specs}

XOR-based coding works on a hop-by-hop basis, i.e. packets encoded by a node are decoded by its neighbors. The idea is that each node $v$ can combine packets using bitwise XOR operations in order to produce an encoded packet. For the neighboring nodes to be able to decode the encoded packet, the choice of native, i.e. non coded, packets is important. More specifically, for a successful coding of $k$ packets , each neighbor should know $k-1$ of those packets beforehand. This requirement guarantees that each neighbor should be able to decode the encoded packet. The existence of $k>1$ packets that can be encoded is known as a \emph{coding opportunity}~\cite{COPE}. It is clear that, finding a coding opportunity depends on $v$'s knowledge about the packets that each of its neighbors has already received. To acquire such information, $v$ employs opportunistic listening~\cite{COPE,CodeB} and snoops all communication in the wireless medium. The acquired information is stored in what is called the \emph{neighbor reception table}. Moreover, node $v$ should store in what is called the \emph{packet pool} all recently received native packets in order to be able to perform decoding of encoded packets. 
To describe the method more formally, let $\mathcal{P}_{v}$ denote $v$'s packet pool, i.e. the set of native packets recently received by $v$ and $\mathcal{R}^u_{v}$ denote $u$'s view of the same buffer. Note that $\mathcal{R}^u_{v}$ is part of $u$'s neighbor reception table. Node $v$ may choose a set of native packets $\mathcal{B}'\!\!\subseteq\!\!\mathcal{P}_{v}$ and produce an encoded packet, by using bitwise XOR, in the presence of a coding opportunity. This means that a set $\mathcal{B}'\!\neq\!\emptyset$,$|\mathcal{B}'|\!>\!1$ can be found such that, according to $v$'s neighbor reception table, each node $u \in \mathcal{N}(v)$ has received at least $|\mathcal{B}'|\!\!-\!\!1$ of the native packets in $\mathcal{B}'$, i.e. $|\mathcal{R}^{v}_{u}\cap \mathcal{B}'|\geq |\mathcal{B}'|\!-\!1, \forall u \in \mathcal{N}(v)$. Successful decoding depends on the consistency of $\mathcal{R}^{v}_{u}$, i.e., whether $\mathcal{R}^{v}_{u}\!\!\subseteq\!\!\mathcal{P}_{u}$. Decoding failures at a node $u$ occur when $|\mathcal{P}_{u}\cap \mathcal{B}'|\!<\!|\mathcal{B}'|\!-\!1$ and result in the loss of all packets included in the encoded one.
	
The efficiency of XOR-based coding clearly depends on the existence of coding opportunities. This is because for each encoded packet that contains $k$ native ones only one transmission is required instead of $k$, thus saving energy and reducing packet collisions. To maximize the number of coding opportunities XOR-based coding approaches introduce a \emph{Random Assessment Delay (RAD)} before relaying a packet. Higher values of RAD result in more candidate packets for encoding, however this comes at the cost of increased end-to-end delay. 

\section{The synergy between network coding and the termination criterion}\label{sec:synergy}
XOR network coding as well as the termination criterion of a CDS-based algorithm are essential mechanisms for energy efficient broadcasting as both aim to minimize packet transmissions. Ensuring a smooth synergy is critical for building an efficient algorithm. Most proposed coding-based broadcast algorithms combine XOR-coding with CDS-based approaches that utilize the M/U criterion~\cite{CodeB,NCDS_conf,NCDS_journal,kunz-iwcmc} while others do not provide insight on the termination criterion used~\cite{kunz-iwcmc,OstovariXOR}. In the following we will show that M/U faces performance issues that hamper the coding operation. At the same time, using other proposed termination criteria, such as R/U and C/U, in parallel with XOR coding raises significant design issues. 

\subsection{The M/U criterion limits coding gains}\label{sec:motivation}
The choice to combine M/U with XOR coding is reasonable. First, M/U is compatible with the RAD technique that is essential for network coding. In fact, RAD improves the pruning efficiency of M/U. This is because the imposed delay allows the reception of more packet duplicates which could potentially change the initial decision to relay the packet. This is the reason for which RAD has also been proposed in the context of broadcasting without network coding~\cite{survey-Williams,survey-Ruiz}. However, although in non-coded approaches there are alternatives that provide performance improvements similar to that of RAD but without the associated delay, in coding-based approaches using RAD is essential. The second reason for which state-of-the-art XOR-based approaches adopt the M/U criterion is because in this case XOR coding can be implemented with limited cost. Recall that the latter requires information about the reception status of the neighboring nodes, i.e. the neighbor reception table. This is exactly the information on which M/U decisions are based on, therefore this information is already available through the implementation of M/U. Besides the advantages of using M/U there is also a significant downside. M/U is known to be less efficient than other proposed criteria~\cite{TC-CU,TC_MRU}. This motivated us to further examine the performance of XOR-based broadcasting implementing M/U against non coding schemes utilizing the other termination criteria, i.e. R/U and C/U. 

\begin{figure}[!t]
	\centering
	{
		\subfloat[Case I][]
		{
			\includegraphics[]{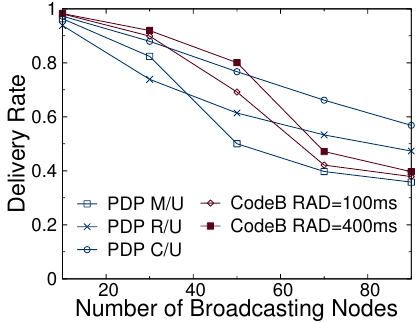}
			\label{motivation_traffic_load_DR}
		}
		\hfil
		\subfloat[Case II][]
		{
			\includegraphics[]{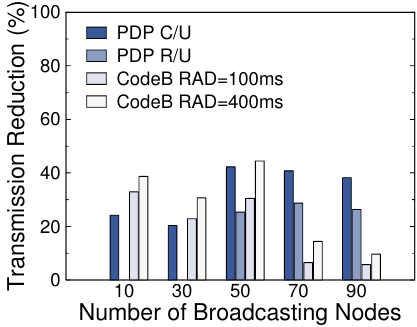}
			\label{motivation_traffic_load_pruning_efficiency}
		}
	}
	\vspace{-10pt}
	\caption{Performance under varying traffic load: (a) Delivery rate (b) Transmission reduction compared to PDP M/U.}
	\label{motivation_traffic_load_plots}
	\vspace{-16pt}
\end{figure}

For our investigation, we conducted a series of experiments using the ns2 simulator~\cite{ns2}. We chose to experiment with the well-established CodeB algorithm~\cite{CodeB} that utilizes network coding and builds on top of PDP using the M/U termination criterion. In our experimental setup, $100$ nodes move with maximum speed of $1$ m/s in a square area according to the Random Waypoint (RW) model~\cite{rwp-dist}. Each node has a neighborhood with an average size of $15$ nodes, while $50$ nodes generate broadcast traffic with a rate of $1$ packet/s. More information about the simulation set-up can be found in Section~\ref{sec:Evaluation}.
First, we evaluated the performance of CodeB under various levels of offered traffic by varying the number of source nodes. Fig.~\ref{motivation_traffic_load_DR} depicts its delivery efficiency, while Fig.~\ref{motivation_traffic_load_pruning_efficiency} displays its ability to prune transmissions compared to the non-coding PDP scheme that uses the M/U criterion. As expected, CodeB successfully reduces the number of transmissions and provides enhanced packet delivery in cases of low to medium traffic load. However, as the network traffic increases its efficiency deteriorates. Based on this observation, we attempted to boost CodeB's performance. More specifically, we included a new version of CodeB that maximizes the benefits of network coding by increasing the RAD value from $200$ ms to $400$ ms. However, despite the fact that the new CodeB version clearly discovers more coding opportunities and thus further reduces transmissions  (Fig.~\ref{motivation_traffic_load_pruning_efficiency}), its improvement in terms of delivery efficiency is limited (Fig.~\ref{motivation_traffic_load_DR}). To further investigate the reasons behind CodeB's poor behavior, we compared it against two versions of the non-coding PDP scheme, one implementing the R/U criterion and the other the C/U. If there exists at least one simple PDP scheme that performs better than CodeB then the origins of the witnessed poor behavior reside in the termination criterion rather than the coding mechanism itself. Interestingly, this is confirmed by our results in Fig.~\ref{motivation_traffic_load_plots}. After the breaking point of $50$ broadcast sources (half of the network nodes), the non-coding PDP schemes outperform CodeB regardless of the RAD value used. The only exception is PDP M/U that, similar to CodeB, suffers from a performance breakdown. 

\begin{figure}[!t]
	\centering
	{
		\subfloat[Case I][]
		{
			\includegraphics[]{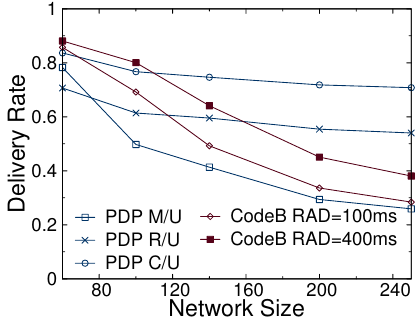}
			\label{motivation_scalability_DR}
		}
		\hfil
		\subfloat[Case II][]
		{
			\includegraphics[]{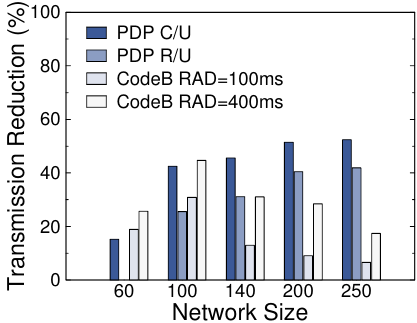}
			\label{motivation_scalability_pruning_efficiency}
		}
	}
	\vspace{-10pt}
	\caption{Performance for different network sizes: (a) Delivery rate (b) Transmission reduction compared to PDP M/U.}
	\label{motivation_scalability_plots}
	\vspace{-12pt}
\end{figure}

Similar findings are witnessed in our second experiment where we assess the scalability of all algorithms by increasing the number of network participants (Fig.~\ref{motivation_scalability_plots}). CodeB outperforms all schemes for networks comprised of fewer than $100$ participants. Despite the fact that the offered load remains constant as the network size increases, CodeB and PDP M/U cannot avoid performance breakdown (Fig.~\ref{motivation_scalability_DR}). Both generate a large number of transmissions (Fig.~\ref{motivation_scalability_pruning_efficiency}) that induce failures due to packet collisions. Increasing the RAD value offers CodeB a performance improvement, however the gain is still limited and the problem is not solved. On the other hand, the non-coding schemes, PDP R/U and C/U, present a relatively stable behavior regardless of the network size. Clearly, the best performing algorithm is PDP C/U that reduces transmissions by up to $\sim$$60\%$ while keeping the delivery efficiency above $\sim$$75\%$.

Overall, the results revealed that the combination of network coding with the M/U termination criterion is not always the best choice. In particular, non-coding schemes perform far better than CodeB when the traffic in the network increases; either because more traffic is offered from more sources (first experiment) or because in a bigger network (second experiment) more forwarders exist and produce more packet duplicates. As explained, this behavior is not the result of the coding operation itself but is inherited from the underlying broadcast scheme and more specifically the termination criterion. There are two reasons for this. The first and predominant one is the limited ability of M/U to prune redundant transmissions. As a result, congestion quickly builds up and results in more collisions, therefore reducing delivery efficiency. The second reason is related to the neighbor reception table, i.e. the structure containing information about the packets received by each neighbor, which is necessary for both M/U and XOR coding. In the typical implementation of this structure, information for each packet is maintained for a limited time period. As traffic in the network increases, the delay jitter between the first and the last duplicate of a packet also increases. As a result, there is an increased probability that a packet duplicate arrives at a node $v$ after the information for that packet has expired and been removed from the neighbor reception table. This results in node $v$ transmitting more duplicates and thus aggravating congestion. Increasing the expiration period for information in the neighbor reception table improves performance up to a limit. After that, no further improvement is possible and performance breakdown is still evident due to the limited pruning ability of M/U. We also implemented the neighbor reception table as a fixed size structure without imposing an expiration period. We tested different sizes and found that performance degradation appears to be more severe in this case.

\subsection{The pitfalls of using other termination criteria}\label{sec:reordering_description}

Our observations highlight the need for replacing M/U with alternative criteria, such as R/U and C/U. However, doing so is not straightforward. The main reason is what we call ``the packet reordering problem''.
Before analyzing the packet reordering problem, let us first describe the implementation aspects of the two alternative termination criteria. Recall that both R/U and C/U delineate a policy for handling packet duplicates. More specifically, in R/U a forwarder relays a duplicate only if no other duplicate of the same packet was relayed in the past. On the other hand, in C/U a forwarder $v$ relays only the packets seen for the first time and ignores packets seen in the past even if $v$ did not forward those packets, i.e. $v$ was not selected as a forwarder at that time. In order for both R/U and C/U to function properly, there are two prerequisites. The first is that packets should be uniquely identified through a number added by the source node at creation time. The second prerequisite is that each node implementing the termination criterion should store a full reception history on a packet basis (i.e. the id's of received packets). This is neither practical nor realistic due to the high storage and processing requirements. For this reason, the traditional implementation of both R/U and C/U takes a much simpler approach. The numbers used to identify packets are assigned by the source in a sequential manner (thus called sequence numbers) so as packets with higher numbers correspond to the ones created more recently. This allows each node $v$ that implements either R/U or C/U to only store a single sequence number ($\mathtt{SN_{s}}$) for every source node $s$. In the  R/U criterion (C/U criterion), this is the largest number seen in a packet from $s$ and forwarded (received) by $v$. Then, for an incoming packet $p_{1}$ carrying the sequence number $\mathtt{SN_{p_{1}}}$ it is sufficient to check that $\mathtt{SN_{p_{1}}}>\mathtt{SN_{s}}$ so as to decide that it is not a duplicate. 

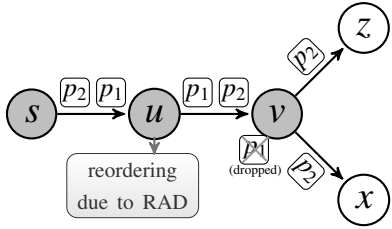
\begin{figure}[!t]
	\centering
	{

	\begin{tikzpicture}[scale=0.95,->,>=stealth',shorten >=1pt,auto,node distance=3cm,
		thick,main node/.style={circle,fill=beaublue,draw,font=\sffamily\Large\bfseries},block/.style={draw,fill=white,rectangle,minimum height=0.58cm,minimum width={width("time instance")}}]
			
		\node[main node,fill=lightgray] (5) at (-3.4,0) {$s$};
		\node[main node,fill=lightgray] (1) at (-1.7,0) {$u$};
		\node[main node,fill=lightgray] (2) at (0,0) {$v$};
		\node[main node,fill=white] (3) at (1.2,1.2) {$z$};
		\node[main node,fill=white] (4) at (1.2,-1.2) {$x$};
			
		\fill [draw,thin,fill=white,rounded corners=2pt] (-3,0.1) rectangle (-2.6,0.5);
		\node[] (p2b) at (-2.8,0.3) {$p_2$};
		\fill [draw,thin,fill=white,rounded corners=2pt] (-2.5,0.1) rectangle (-2.1,0.5);
		\node[] (p1b) at (-2.3,0.3) {$p_1$};
			
		\fill [draw,thin,fill=white,rounded corners=2pt] (-1.3,0.1) rectangle (-0.9,0.5);
		\node[] (p1) at (-1.1,0.3) {$p_1$};
		\fill [draw,thin,fill=white,rounded corners=2pt] (-0.8,0.1) rectangle (-0.4,0.5);
		\node[] (p2) at (-0.6,0.3) {$p_2$};
			
		\fill [draw,thin,fill=white,rounded corners=2pt,rotate around={45:(0.4,0.8)}] (0.2,0.6) rectangle (0.6,1.0);
		\node[rotate=45] (pp2) at (0.4,0.8) {$p_2$};
			
		\fill [draw,thin,fill=white,rounded corners=2pt,rotate around={-45:(0.4,-0.8)}] (0.2,-0.6) rectangle (0.6,-1.0);
		\node[rotate=-45] (pp2) at (0.4,-0.8) {$p_2$};
			
		\node[shape=rectangle,ultra thin,black!80,text width=1.5cm,align=center,rounded corners,draw,top color=white, bottom color=black!10] (reordering) at (-2,-1) {\footnotesize reordering due to RAD};
			
		\path[every node/.style={font=\sffamily\small},->,color=black!55]
		(-1.7,-0.32) edge [bend left=0] node {} (-1.7,-0.61);
			
		\fill [draw,thin,fill=white,rounded corners=2pt] (-0.52,-0.7) rectangle (-0.12,-0.3);
		\node[] (fp1) at (-0.3,-0.5) {$p_1$};
		\node[] (drop) at (-0.3,-0.8) {\tiny (dropped)};
			
		\fill [draw,-] (-0.3,-0.5) node[cross=0.2cm,rotate=0,gray!80]{};
			
		\path[every node/.style={font=\sffamily\small}]
			(5) edge node [above] {} (1)
			(1) edge node [above] {} (2)
			(2) edge node [above] {} (3)
			(2) edge node [above] {} (4);
			
			
			
			
			
	\end{tikzpicture}
	}
	\caption{Example where the propagation of packet $p_1$ terminates due to the packet reordering problem.}
	\label{example_CU_failure}
	\vspace{-12pt}
\end{figure}

Unfortunately, the aforementioned implementation is fully functional only under the assumption that all nodes in the network receive packets in the same order in which they were created. When this order is altered the problem that we call \emph{packet reordering} emerges, impairing the ability of both  R/U and C/U to detect duplicates and therefore having a severe impact on their performance. The utilization of XOR coding unfortunately results in packet reordering and thus its incompatibility with current implementations of both R/U and C/U. More specifically, packet reordering appears due to the random assessment delay (RAD) that network coding uses at each node in order to maximize the probability of finding a coding opportunity. To make it more clear, let us examine the problem through an example in which both R/U and C/U fail to work properly. 
Fig.~\ref{example_CU_failure} illustrates the propagation of packets $p_1$ and $p_2$ across an example network. Both packets originate from the same source $s$ and $p_1$ is created before $p_2$. Therefore, the sequence number of $p_1$ is smaller than that of $p_2$, i.e., $\mathtt{SN_{p_{1}}}\!\!<\!\mathtt{SN_{p_{2}}}$. 
When RAD is not utilized, $u$ will forward both packets in the same order as received. Then, $v$ will first receive $p_{1}$, update $\mathtt{SN_{s}}$, i.e. $\mathtt{SN_{s}}\leftarrow\mathtt{SN_{p_{1}}}$, and finally forward $p_{1}$. Upon reception of $p_{2}$, $v$ will confirm that $\mathtt{SN_{p_{2}}}\!\!>\!\!\mathtt{SN_{s}}$ and will forward $p_{2}$. 
On the other hand, if RAD is utilized, $u$ introduces a random delay before forwarding $p_{1}$ and $p_{2}$. Due to randomness, the delay for $p_{2}$ may be significantly smaller than the corresponding delay for $p_{1}$, thus resulting in $u$ forwarding the two packets in the reverse order, i.e. $p_{2}$ first and then $p_{1}$. After receiving $p_{2}$, $v$ updates $\mathtt{SN_{s}}$ to the value $\mathtt{SN_{p_{2}}}$ and forwards $p_{2}$. Later on, when $v$ receives $p_{1}$ makes the observation that $\mathtt{SN_{p_{1}}}\!\!<\!\!\mathtt{SN_{s}}$, therefore rejects $p_{1}$ although it is not a duplicate. This decision has a major impact on the delivery efficiency since $p_{1}$ never reaches nodes $z$ and $x$.

\begin{figure}[!t]
	\centering
	{
		\subfloat[Case I][]
		{
			\includegraphics[]{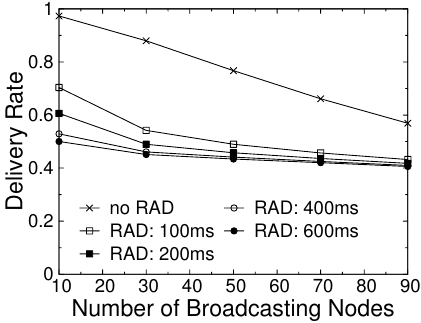}
			\label{PDP_CU_traffic_load_DR}
		}
		\subfloat[Case II][]
		{
			\includegraphics[]{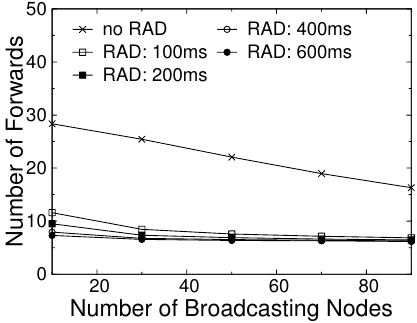}
			\label{PDP_CU_traffic_load_FW}
		}
	}
		\vspace{-6pt}
	\caption{Performance of PDP C/U vs traffic load using different RAD intervals: (a) Delivery efficiency (b) Average number of forwards per packet.}
	\label{PDP_CU_ttl_traffic_load_plots}
	\vspace{-12pt}
\end{figure}

To validate the impact of the packet reordering problem we conducted a series of experiments on the PDP algorithm using the C/U termination criterion. We examined the effect of different values of the random assessment delay. For the experimental setup we used the same settings described in Section~\ref{sec:motivation}. Fig.~\ref{PDP_CU_ttl_traffic_load_plots} illustrates our main results. Clearly, RAD has a considerable impact on the overall broadcasting performance regardless of the offered load. More specifically, the packet reordering problem may result in a reduction of the number of relaying decisions up to $\sim$$3$ times (Fig.~\ref{PDP_CU_traffic_load_FW}). However, this pruning is erroneous in the sense that it prematurely terminates the broadcasting process, thus, reducing the delivery efficiency up to $\sim$$50\%$ (Fig.~\ref{PDP_CU_traffic_load_DR}). As expected, higher RAD values have a more severe impact on the performance as in these cases the probability of receiving packets out of order increases. We observed similar findings in the R/U case.

\section{Building a coding friendly termination criterion}\label{sec:MCU}
Establishing the compatibility of R/U and C/U with XOR coding requires solving the packet reordering problem. Allowing each node that implements either of these criteria to store a full packet reception history can provide a solution. However, as mentioned earlier, this approach is neither practical nor realistic due to the high storage and processing requirements. 
Towards a more efficient solution, we propose the \emph{modified covered/uncovered (MC/U)} termination criterion that extends C/U. We choose to build on top of C/U because both the related literature~\cite{TC-CU,TC_MRU} and our experimental results (Fig.~\ref{motivation_traffic_load_plots} and Fig.~\ref{motivation_scalability_plots}) confirm that it achieves the best performance against all other proposed termination criteria.
The main idea behind our approach is to implement the same forwarding criteria as in C/U but to allow each node to store information (just one bit as we will discuss in the following) for each of the $k$, instead of just one, most recently seen packets from each source node. 
This allows the node to detect any duplicate of these $k$ packets without problems caused by packet reordering. Duplicate detection is not possible for a packet that is older than the $k$ recorded ones because no relevant information is available. However, this is important only if a copy of a packet $p$ from source $s$ is received by a node after the $k$-th packet that $s$ generated after $p$. By increasing $k$ it is possible to minimize the probability of such an occasion. Even if such an occasion arises we choose to drop the packet, i.e. adopt the C/U policy, rather than forwarding it (which corresponds to the M/U policy) in order to avoid increasing the network congestion levels.

\begin{figure}[!t]
	\centering
	{
		\begin{tikzpicture}
		\foreach \c/\i [count=\n] in {white/,white/,lightgray/,white/,white/\dots,white/,white/} 
		\node[draw,fill=\c,minimum height=0.5cm,minimum width=0.7cm](A\n) at (\n*0.7cm,0) {\i} ;
		\foreach \c/\i [count=\n] in {white/,white/,white/,white/,white/\dots,lightgray/,white/} 
		\node[draw,fill=\c,minimum height=0.5cm,minimum width=0.7cm](B\n) at (0.2cm+\n*0.7cm,-0.2) {\i} ;
		\foreach \c/\i [count=\n] in {white/,white/,white/,lightgray/,white/\dots,white/,white/} 
		\node[draw,fill=\c,minimum height=0.5cm,minimum width=0.7cm](C\n) at (0.4cm+\n*0.7cm,-0.4) {\i} ;
		\foreach \c/\i [count=\n] in {white/,lightgray/,white/,white/,white/\dots,white/,white/} 
		\node[draw,fill=\c,minimum height=0.5cm,minimum width=0.7cm](D\n) at (0.6cm+\n*0.7cm,-0.6) {\i} ;
		
		\fill [draw,-,color=black,fill=white,decorate,decoration=brace,xshift=-1em,yshift=-2em](0.6,1) -- (5.6,1) node [black,midway,yshift=0.3cm,text width=2.5cm,align=center]{\footnotesize $k$};
		
		\fill [draw,-,color=black,fill=white,decorate,decoration=brace,xshift=-0.9em,yshift=-2em,right](5.8,1) -- (6.5,-0.2) node [black,midway,yshift=0.1cm, xshift=-0.4cm, text width=3cm,align=center]{\footnotesize one bitmap per source node};
		
		\node[] (SN) at (7,-1.3) {\footnotesize maximum sequence number};
		
		\path[every node/.style={font=\sffamily\small},->]
		(SN.west) edge [bend left=20] node {} (D2.center)
		(SN.west) edge [bend left=30] node {} (3.2,-0.25)
		(SN.west) edge [bend left=35] node {} (4.4,-0.05)
		(SN.west) edge [bend left=30] node {} (2.1,0.175);
		
		\end{tikzpicture}
	}
	\vspace{-6pt}
	\caption{Structure used for the modified covered/uncovered (MC/U) termination criterion.}
	\label{CU_modified}
	\vspace{-12pt}	
\end{figure}
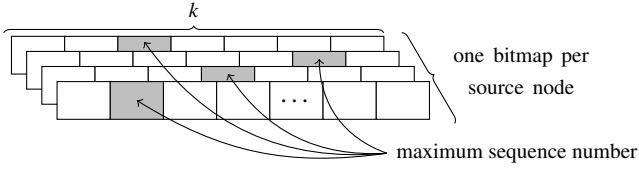
\begin{figure}[!b]
	\vspace{-12pt}
	\begin{center}
		\line(1,0){250}
	\end{center}
	{
		\footnotesize
		\vspace{-14pt}
		$\;$\textbf{RelayOrNot(packet $p$, bitmap $\mathtt{BM_{s}}$, int $\mathtt{SN^{MAX}_{s}}$, int $\mathtt{mindex}$)}
		\vspace{-17pt}
	}
	\begin{center}
		\line(1,0){250}
	\end{center}
	\vspace{-8pt}
	\begin{algorithmic}[1]
		\footnotesize
		
		\IF{($p.\mathtt{SN}>\mathtt{SN^{MAX}_{s}}$)}
		\STATE Update(p,$\mathtt{BM_{s}}$,$\mathtt{SN^{MAX}_{s}}$,$\mathtt{mindex}$)
		\STATE \textbf{relay} $p$ \textbf{if forwarder}
		\ELSE
		\STATE $\mathtt{SN_{s}^{MIN}}\leftarrow\mathtt{SN_{s}^{MAX}}-k$
		\IF{($p.\mathtt{SN} <= \mathtt{SN_{s}^{MIN}}$)}
		\STATE \textbf{drop} $p$
		\ELSE
		
		\STATE $\mathtt{index}\leftarrow p.\mathtt{SN}-\mathtt{SN_{s}^{MAX}}+\mathtt{mindex}$
		\IF{($\mathtt{index} < 0$)}
		\STATE $\mathtt{index}\leftarrow\mathtt{index}+k$
		\ENDIF
		\STATE $val\leftarrow\mathtt{BM_{s}}.get(\mathtt{index})$
		
		\IF{($val$)}
		\STATE \textbf{drop} $p$
		\ELSE
		\STATE $\mathtt{BM_{s}}.set(\mathtt{index})$ 
		\STATE \textbf{relay} $p$ \textbf{if forwarder}
		\ENDIF
		\ENDIF
		\ENDIF
		
		\normalsize
	\end{algorithmic}
	\vspace{-20pt}
	\begin{center}
		\line(1,0){250}
	\end{center}
	\vspace{-16pt}
	\caption{Pseudocode of the MC/U forwarding procedure.}
	\label{MCU_pseudocode}
\end{figure}

Selecting a proper value for $k$ is clearly a challenging task. Large values increase the storage and processing requirements at each node while small values increase the probability of receiving a packet without being able to decide whether it is a duplicate or not. After experimentation, we concluded that the MC/U criterion has a competitive performance even when a small value of $k$ is required due to space limitations. Nonetheless, the storage and processing requirements may raise a concern. 
To address such concerns, we implement MC/U using bitmaps. Note that bitmaps have been used for similar purposes in the context of multicasting in ad hoc networks~\cite{Jetcheva}. More specifically, a node $v$ that implements MC/U, instead of storing the $k$ last seen sequence numbers from a source $s$, it uses only one bit for each of them, i.e. a total of $k$ bits in the form of a bitmap $\mathtt{BM_{s}}$ (Fig.~\ref{CU_modified}). Then sequence numbers are mapped to the bits of $\mathtt{BM_{s}}$ and each bit is used to indicate whether the corresponding sequence number is known (bit set to 1), i.e. $v$ has already received a packet carrying this sequence number, or not (bit set to 0).
At the same time, by using bitmaps the node takes advantage of the low cost read/write operations. Furthermore, node $v$ stores the maximum known sequence number from $s$ ($\mathtt{SN_{s}^{MAX}}$) as well as the index ($\mathtt{mindex}$) of the bit in $\mathtt{BM_{s}}$ that corresponds to $\mathtt{SN_{s}^{MAX}}$.

The pseudocode of MC/U is illustrated in Fig.~\ref{MCU_pseudocode}. When a node $v$	receives a packet $p$ from $s$ it first checks whether its sequence number $p.\mathtt{SN}$ is greater than $\mathtt{SN_{s}^{MAX}}$. If this is the case $v$ relays the packet (if it is an elected forwarder) and updates its state (Fig.~\ref{Update_pseudocode}). This update involves the following steps. First $v$ calculates the index ($\mathtt{mindex'}$) of the bit that corresponds to the new sequence number (line 1). Observe that this calculation may involve a rollover, i.e. reusing the bits of the bitmap. 
Then $v$ resets all the bits from position $\mathtt{mindex}\!+\!1$ to $\mathtt{mindex'}\!\!-\!1$ (lines 2-10). This is done because those bits correspond to the sequence numbers between $\mathtt{SN_{s}^{MAX}}$ and $p.\mathtt{SN}$ and no packet carrying one of these numbers has been received so far. Note that if $p.\mathtt{SN}-\mathtt{SN_{s}^{MAX}}\!>\!k$, i.e. a multiple rollover occurs, then all bits of the bitmap must be reset (lines 6-7). Finally, $v$ updates $\mathtt{SN_{s}^{MAX}}$ and $\mathtt{mindex}$ (lines 11-12) and sets the corresponding bit to indicate that a packet carrying $\mathtt{SN_{s}^{MAX}}$ has already been received (line 13). 
Going back to the basic algorithm (Fig.~\ref{MCU_pseudocode}), if $p.\mathtt{SN}\leq\mathtt{SN_{s}^{MAX}}$ then $v$ should decide whether $p.\mathtt{SN}$ is one of the $k$ most recent sequence numbers. If not (lines 5-7) the packet is dropped because it is not possible to decide whether it is a duplicate or not. Otherwise, $v$ calculates the index of the bit that corresponds to $p.\mathtt{SN}$ (lines 9-12). If that bit is set to 1 then $p$ is dropped because it is a duplicate otherwise the bit is set to 1 and $p$ is relayed if $v$ is an elected forwarder (lines 13-18).

\begin{figure}[!t]
	\begin{center}
		\line(1,0){250}
	\end{center}
	{
		\footnotesize
		\vspace{-14pt}
		$\;$\textbf{Update(packet $p$, bitmap $\mathtt{BM_{s}}$, int $\mathtt{SN^{MAX}_{s}}$, int $\mathtt{mindex}$)}
		\vspace{-17pt}
	}
	\begin{center}
		\line(1,0){250}
	\end{center}
	\vspace{-8pt}
	\begin{algorithmic}[1]
		\footnotesize
		\STATE $\mathtt{mindex'} \leftarrow (\mathtt{mindex}+p.\mathtt{SN}-\mathtt{SN^{MAX}_{s}})\%k$
		\STATE $\mathtt{rollover} \leftarrow \Big\lfloor\frac{\mathtt{mindex}+p.\mathtt{SN}-\mathtt{SN^{MAX}_{s}}}{k}\Big\rfloor$
		\IF{($\mathtt{rollover} == 1$)}
		\STATE $\mathtt{BM_{s}}.zero(\mathtt{mindex}+1,k-1)$
		\STATE $\mathtt{BM_{s}}.zero(0,\mathtt{mindex'}-1)$
		\ELSIF{($\mathtt{rollover} > 1$)}
		\STATE $\mathtt{BM_{s}}.zero(0,k-1)$
		\ELSE
		\STATE $\mathtt{BM_{s}}.zero(\mathtt{mindex}+1,\mathtt{mindex'}-1)$
		\ENDIF
		\STATE $\mathtt{mindex} \leftarrow \mathtt{mindex'}$
		\STATE $\mathtt{SN^{MAX}_{s}} \leftarrow p.\mathtt{SN}$
		\STATE $\mathtt{BM_{s}}.set(\mathtt{mindex})$
		\normalsize
	\end{algorithmic}
	\vspace{-20pt}
	\begin{center}
		\line(1,0){250}
	\end{center}
	\vspace{-16pt}
	\caption{Pseudocode for updating the bitmap.}
	\label{Update_pseudocode}
	\vspace{6pt}
\end{figure}

\begin{figure}[!t]
	\vspace{-16pt}
	\centering
	{
		\subfloat[Case I][]
		{
			\includegraphics[]{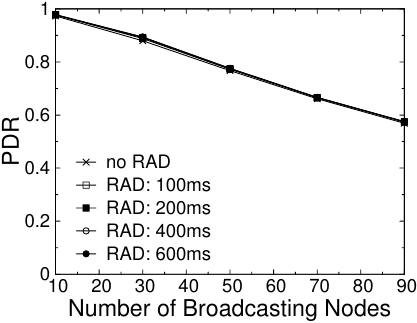}
			\label{PDP_MCU_traffic_load_DR}
		}
		\subfloat[Case II][]
		{
			\includegraphics[]{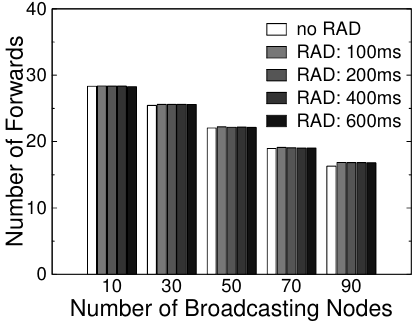}
			\label{PDP_MCU_traffic_load_Delay}
		}
	}
	\vspace{-12pt}
	\caption{Performance of PDP MC/U under varying traffic load using different RAD intervals: (a) Delivery efficiency (b) Average number of forwards per packet.}
	\label{PDP_MCU_ttl_traffic_load_plots}
	\vspace{-12pt}
\end{figure}
To validate the efficacy of MC/U we developed a version of the PDP scheme that utilizes it instead of C/U. Then, we repeated the same experiments described in Section~\ref{sec:reordering_description}, testing different values of RAD under varying offered load. Fig.~\ref{PDP_MCU_ttl_traffic_load_plots} illustrates our main results. In contrast to PDP C/U (Fig.~\ref{PDP_CU_traffic_load_DR}), RAD has a negligible impact on the delivery performance of PDP MC/U (Fig.~\ref{PDP_MCU_traffic_load_DR}). At the same time, the ability of MC/U to prune transmissions is not damaged. Both MC/U and C/U achieve roughly the same number of transmissions (compare Fig.~\ref{PDP_MCU_traffic_load_Delay} and \ref{PDP_CU_traffic_load_FW} when the performance of C/U does not collapse, i.e. when no RAD is used). Overall, the results prove that MC/U successfully tackles the packet reordering problem.

\section{Network Coding Broadcast with Coded Redundancy}

In this section, we introduce the \emph{Network cOding Broadcast with Coded Redundancy} (NOB-CR) algorithm. NOB-CR, similar to other schemes, takes the approach to implement XOR coding on top of a CDS-based broadcast algorithm. However, in order to maximize the performance of network coding, NOB-CR employs the MC/U termination criterion introduced in section~\ref{sec:MCU}. As the default CDS algorithm we choose PDP although any algorithm of this category could be used. Regarding the coding process, similar to XOR coding approaches, NOB-CR utilizes network coding on a hop-by-hop basis. Each intermediate node uses bitwise XOR operations to combine native packets into encoded ones under the requirement that all neighboring nodes can decode them. Nonetheless, NOB-CR uses only one of the two specialized data structures required for coding (see Section~\ref{sec:coding_specs}), i.e. the packet pool. This is because its lightweight coding detection mechanism renders obsolete the use of the other one, i.e. the neighbor reception table. We discuss this issue in detail in Section~\ref{sec:coding_opp}. Besides the aforementioned differences, NOB-CR deviates from other approaches by introducing a series of mechanisms that significantly improve the broadcasting performance and alleviate the related costs. In particular, these mechanisms are:
\begin{itemize}
	\item A lightweight coding detection method that operates without the need of maintaining a neighbor reception table.
	\item A novel method for the computation of forwarders that uses information provided by the RAD mechanism to increase the overall pruning efficiency.
	\item A cost-free method to inject packet redundancy in the network in order to reduce the end-to-end delay.
\end{itemize}
In the following, we delineate NOB-CR's basic operation as well as the aforementioned mechanisms.

\subsection{Basic operation}
\begin{figure}[!t]
	\centering
	{
		\begin{tikzpicture}[%
		>=triangle 60,              
		start chain=going below,   
		node distance=6mm and 60mm,
		every join/.style={norm},
		]
		
		\tikzset{
			base/.style={draw, on chain, on grid, align=center, minimum height=4ex},
			proc/.style={base, rectangle, text width=6em, scale=0.7},
			test/.style={base, diamond, aspect=1.6, text width=4em,scale=0.7},
			term/.style={proc, rounded corners},
			coord/.style={coordinate, on chain, on grid, node distance=6mm and 25mm},
			nmark/.style={draw, cyan, circle, font={\sffamily\bfseries}},
			norm/.style={->, draw},
			free/.style={->, draw},
			cong/.style={->, draw},
			it/.style={font={\small\itshape}}
		}
		\node [proc, fill=lightgray] (p0) {packet reception};
		\node [test,right=2.2cm of p0, join] (t1) {Encoded?};
		\node [proc, below=1.2cm of t1] (p1) {Decode};
		\node [test, right=2.9cm of t1,yshift=-1cm] (t3) {MC/U fulfilled?};
		\node [proc, above=1cm of t3] (p2) {Drop packet};
		\node [test, below=1.5cm of t3] (t4) {Forwarder?};
		\node [test, below=1.5cm of t4] (t5) {Coding exists?};
		\node [proc, left=2.2cm of t5,yshift=-1.2cm,xshift=0.6cm] (p3) {Buffered for RAD};
		\node [test, left=1.9cm of p3, scale=0.6, aspect=1,text width=6em] (t6) {{\Large RAD expired?}};
		\node [test, below=3cm of p0, aspect=1,xshift=0.3em] (t7) {Coding exists?};
		\node [proc, right=2.6cm of t7] (p4) {Compute fwds $\forall$ native pkt};
		\node [proc, above=1.5cm of t7] (p5) {Compute fwds};
		\node [proc, above=0.8cm of p5] (p6) {Send as native};
		\node [proc, above=1cm of p4] (p7) {Send encoded};

		\node [coord, right=1.3cm of t1] (c0)  {};
		\node [coord, right=1.3cm of p1] (c1)  {};
		\node [coord, left=1.6cm of t3] (c2)  {};
		\path (t1.south) to node [near start, xshift=0.8em] {{\small $yes$}} (p1);
		\draw [->] (t1.south) -- (p1);
		\draw [-] (p1.east) -- (c1);
		\draw [-] (c1) -- (c2);
		
		\path (t1.east) to node [near start, yshift=0.7em,xshift=-0.4em] {{\small $no$}} (t3);
		\draw [-] (t1.east) -- (c0);
		\draw [-] (c0) -- (c2);
		\draw [->] (c2) -- (t3);

		\path (t3.south) to node [near start, xshift=1em] {{\small $yes$}} (t4);
		\draw [->] (t3.south) -- (t4);
		
		\node [coord, right=1.5cm of t3] (c6) {};
		\path (t3.east) to node [near start, yshift=0.5em] {{\small $no$}} (c6);
		\node [coord, below=1.5cm of c6] (c7) {};
		\path (t4.east) to node [near start, yshift=0.5em] {{\small $no$}} (c7);
		\node [coord, above=1cm of c6] (c8) {};
		\draw [-] (c6.south) -- (c7);
		\draw [-] (t4.east) -- (c7);
		\draw [-] (t3.east) -- (c6);
		\draw [-] (c6) -- (c8);
		\draw [->] (c8) -- (p2);
		
		\path (t4.south) to node [near start,xshift=1em] {{\small $yes$}} (t5);
		\draw [->] (t4.south) -- (t5);
		
		\node [coord, below=0.85cm of t5] (c5) {};
		\path (t5.south) to node [near start, yshift=0em,xshift=0.8em] {{\small $no$}} (c5);
		\draw [-] (t5.south) -- (c5);
		\draw [->] (c5) -- (p3);
		
		\draw [->] (p3.west) -- (t6);
		\node [coord, below=0.9cm of t6] (c9) {};
		\path (t6.south) to node [near start,xshift=0.8em] {{\small $no$}} (c9);
		\draw [-] (t6.south) -- (c9);
		\node [coord, below=0.9cm of p3] (c10) {};
		\draw [-] (c9.east) -- (c10);
		\draw [->] (c10.north) -- (p3);
		
		\node [coord, below=0.7cm of p4] (c15) {};
		\path (t5.west) to node [near start,yshift=0.5em,xshift=0.6em] {{\small $yes$}} (c15);
		\draw [-] (t5.west) -- (c15);
		\draw [->] (c15) -- (p4);
		
		\node [coord, below=1.544cm of t7] (c11) {};
		\path (t6.west) to node [near start,yshift=0.4em,] {{\small $yes$}} (c11);
		\draw [-] (t6.west) -- (c11);
		\draw [->] (c11.east) -- (t7);
		
		\path (t7.east) to node [near start,yshift=0.4em] {{\small $yes$}} (p4);
		\draw [->] (t7.east) -- (p4);
		
		\path (t7.north) to node [near start,xshift=0.6em] {{\small $no$}} (p5);
		\draw [->] (t7.north) -- (p5);
		\draw [->] (p5.north) -- (p6);
		\draw [->] (p4.north) -- (p7);
		
		\end{tikzpicture}
	}
	\vspace{-6pt}
	\caption{Flow diagram of a packet's life cycle in NOB-CR.}
	\vspace{-12pt}
	\label{NOB-CR-flowchart}
\end{figure}
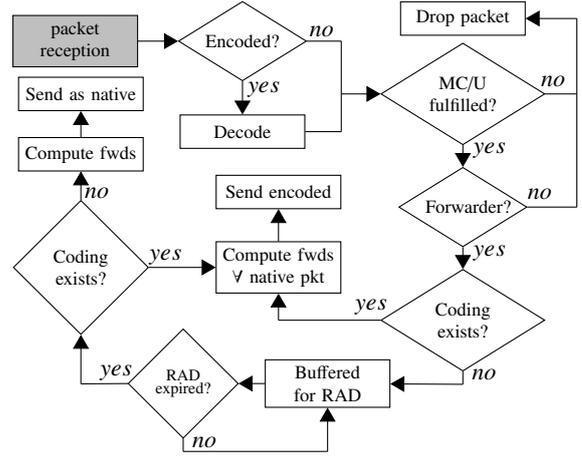
Fig.~\ref{NOB-CR-flowchart} describes the basic operation of NOB-CR by illustrating the life cycle of a packet at a node that implements the protocol. An incoming packet is first examined for deciding if it is a native or an encoded one. In the latter case, the packet is decoded to produce the native packets that it consists of. Then, for each native packet the receiving node examines: a) whether the packet meets the termination criterion conditions, and b) the set of forwarders that is piggybacked on the packet to determine if it is selected as a forwarder for the packet. If at least one of these tests is negative the packet is dropped and the process terminates. Otherwise, the coding opportunity detection process initiates. If a coding opportunity is detected, the set of forwarders is determined (please refer to Section~\ref{sec:coding_specs} for details) for each native packet involved and then the encoded packet is created and immediately transmitted. In the absence of a coding opportunity, the received packet is temporarily buffered in the output queue for a randomly chosen time interval according to the RAD mechanism. This allows the packet to participate in subsequent coding inquiries. When the buffering interval expires the packet is examined for coding one last time before being transmitted as native.

\subsection{Lightweight detection of coding opportunities}\label{sec:coding_opp}
\begin{figure}[!b]
	\vspace{-18pt}
	\centering
	{
		\subfloat[Case I][]
		{
			\begin{tikzpicture}

			\fill[lightgray] \secondcircle;
			
			\begin{scope}
			\clip \firstcircle;
			\clip \secondcircle;
			\clip \thirdcircle;
			\fill[white] \secondcircle;
			\fill[pattern=north west lines, pattern color=lightgray] \secondcircle;
			\end{scope}

			\begin{scope}[even odd rule]
			\clip \thirdcircle (-1.2,-1.2) rectangle (2,2);
			\clip \firstcircle (-1.2,-1.2) rectangle (2,2);
			\fill[white] \secondcircle;
			\fill[pattern=crosshatch, pattern color=lightgray] \secondcircle;
			\end{scope}
			
			\draw[dashed] \firstcircle;
			\draw \secondcircle;
			\draw[dashed] \thirdcircle;

			\filldraw[black,fill=white] (0,0) circle (0.05cm);
			\filldraw[black,fill=white] (0.4,0.7) circle (0.05cm);
			\filldraw[black,fill=white] (0.3,1.4) circle (0.05cm);
			
			\node[] (u) at (-0.15,0) {\textbf{$u$}};
			\node[] (v) at (0.25,0.7) {\textbf{$v$}};
			\node[] (w) at (0.1,1.4) {\textbf{$w$}};
			
			\path[->,thick]
			(0.03,0.04) edge node [above] {} (0.37,0.65)
			(0.31,1.35) edge node [above] {} (0.39,0.75);
			
			\fill [draw,thin,fill=white,rounded corners=2pt] (0.2,0.05) rectangle (0.5,0.35);
			\node[] (p1) at (0.36,0.2) {\small$p_1$};
			
			\fill [draw,thin,fill=white,rounded corners=2pt] (0.4,1.0) rectangle (0.7,1.3);
			\node[] (p2) at (0.55,1.15) {\small$p_2$};
			
			\node[] (R) at (1.8,-0.3) {\textbf{$\mathcal{R}$}};
			\filldraw[white,fill=white] (1.38,0.45) circle (0.01cm);
			\path[every node/.style={font=\sffamily\small},->]
			(1.8,-0.1) edge [bend left=-30] node {} (1.38,0.45);
			
			\node[] (S) at (-1.2,0.85) {\textbf{$\mathcal{S}$}};
			\path[every node/.style={font=\sffamily\small},->]
			(-1.05,0.85) edge [bend left=30] node {} (-0.3,0.8);
			
			\node[] (Nv) at (1.9,1.33) {\textbf{$\mathcal{N}(v)$}};
			\path[every node/.style={font=\sffamily\small},->]
			(1.85,1.2) edge [bend left=30] node {} (1.57,1);
			
			\end{tikzpicture}
			\label{coding_opportunities_2_prev}
		}
		\hfil
		\subfloat[Case II][]
		{
			\begin{tikzpicture}

			\fill[lightgray] \secondcircle;
			
			\begin{scope}
			\clip \firstcircle;
			\clip \secondcircle;
			\clip \thirdcircle;
			\fill[white] \secondcircle;
			\fill[pattern=north west lines, pattern color=lightgray] \secondcircle;
			\end{scope}
			
			\begin{scope}
			\clip \secondcircle;
			\clip \thirdcircle;
			\clip \fourthcircle;
			\fill[white] \secondcircle;
			\fill[pattern=north west lines, pattern color=lightgray] \secondcircle;
			\end{scope}

			\draw[dashed] \firstcircle;
			\draw \secondcircle;
			\draw[dashed] \thirdcircle;
			\draw[dashed] \fourthcircle;

			\filldraw[black,fill=white] (0,0) circle (0.05cm);
			\filldraw[black,fill=white] (0.4,0.7) circle (0.05cm);
			\filldraw[black,fill=white] (0.3,1.4) circle (0.05cm);
			\filldraw[black,fill=white] (1.25,0.45) circle (0.05cm);
			
			\node[] (u) at (-0.15,0) {\textbf{$u$}};
			\node[] (v) at (0.25,0.7) {\textbf{$v$}};
			\node[] (w) at (0.1,1.4) {\textbf{$w$}};
			\node[] (z) at (1.4,0.45) {\textbf{$z$}};
			
			\path[->,thick]
			(0.03,0.04) edge node [above] {} (0.37,0.65)
			(0.31,1.35) edge node [above] {} (0.39,0.75)
			(1.2,0.47) edge node [above] {} (0.46,0.69);
			
			\fill [draw,thin,fill=white,rounded corners=2pt] (0.2,0.05) rectangle (0.5,0.35);
			\node[] (p1) at (0.36,0.2) {\small$p_1$};
			
			\fill [draw,thin,fill=white,rounded corners=2pt] (0.4,1.0) rectangle (0.7,1.3);
			\node[] (p2) at (0.55,1.15) {\small$p_2$};
			
			\fill [draw,thin,fill=white,rounded corners=2pt] (0.8,0.2) rectangle (1.1,0.5);
			\node[] (p1b) at (0.95,0.35) {\small$p_1$};
			
			\node[] (S) at (-1.2,0.85) {\textbf{$\mathcal{S}$}};
			\path[every node/.style={font=\sffamily\small},->]
			(-1.05,0.85) edge [bend left=30] node {} (-0.3,0.8);
			
			\node[] (Nv) at (1.9,1.8) {\textbf{$\mathcal{N}(v)$}};
			\path[every node/.style={font=\sffamily\small},->]
			(1.85,1.67) edge [bend left=30] node {} (1.57,1);
			
			\end{tikzpicture}
			\label{coding_opportunities_3_prev}
		}
		
	}
	\caption{Searching for coding opportunities when $v$ has received (a) one duplicate, or (b) more duplicates per packet.}
	\label{coding_opportunities_example}
\end{figure}
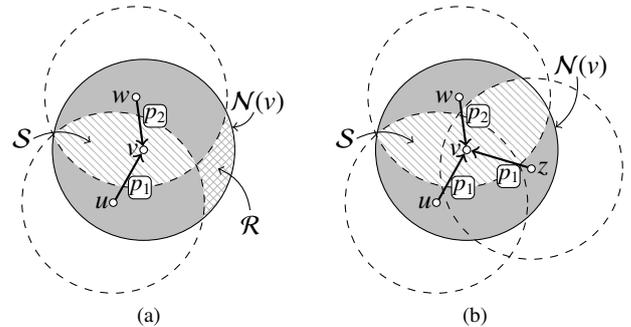
Finding coding opportunities at an intermediate node strongly depends on the packets that each neighbor has already received. To acquire such information, each node snoops all communication in the wireless medium~\cite{COPE,CodeB} and stores it in the neighbor reception table. However, the maintenance of this table comes at a significant cost. Keeping track of every packet received by each neighbor requires a considerable amount of storage. Likewise, updating the neighbor reception table on a packet arrival basis is a task that requires significant processing power. In order to avoid these costs, we introduce a new approach for finding coding opportunities. Our method operates without the need for a neighbor reception table. In fact, it uses neighborhood information that is available through the underlying broadcasting mechanism.
To explain the method, let us use the example in Fig.~\ref{coding_opportunities_2_prev}. In this, node $v$ receives packets $p_1$ and $p_2$ from $u$ and $w$ respectively and checks for a coding opportunity. Recall that a coding opportunity exists only when all of $v$'s neighbors can decode the prospective encoded packet $p_1$$\oplus$$p_2$. In other words, all neighbors must know either $p_1$ or $p_2$, or both. Observe that the set of neighbors that can decode the packet is $\mathcal{N}(u)\cup\mathcal{N}(w)$. As a result, it suffices for $v$ to confirm that the set $\mathcal{R}\!\!=\!\!\mathcal{N}(v)\!-\!\mathcal{N}(u)\!-\!\mathcal{N}(w)$ is an empty set to decide that $p_1$$\oplus$$p_2$ is possible.

To increase the probability of producing an encoded packet our approach takes advantage of packet duplicates. Recall that a native packet waits in node $v$ for a random time in order to find a coding opportunity. During this time $v$ receives multiple copies of the native packet. Our observation is that each of these copies reaches a different part of $v$'s neighborhood. As a result, it is possible to minimize $\mathcal{R}$ and thus increase the probability of finding a coding opportunity. Fig.~\ref{coding_opportunities_3_prev} illustrates the advantages of considering packet duplicates. 
In the example, node $v$ receives a duplicate of $p_{1}$ from node $z$ while initially $v$ received $p_{1}$ from node $u$. Taking into account the neighbors that indirectly received $p_{1}$ through node $z$, node $v$ searches for coding opportunities by estimating the set $\mathcal{R}\!\!=\!\!\mathcal{N}(v)\!-\!\mathcal{N}(w)\!-\!(\mathcal{N}(u)\cup\mathcal{N}(z))$, which is now an empty set and therefore coding is possible. In general, when multiple duplicates of both $p_1$ and $p_2$ exist, $v$ can detect coding opportunities using the set 
\begin{equation}
\mathcal{R}\!\!=\!\!\mathcal{N}(v)\!-\!\mathcal{Z}_{1}\!-\!\mathcal{Z}_{2}
\vspace{-8pt}
\end{equation}
where
\begin{equation}
\mathcal{Z}_{m} = \bigcup\limits_{i\in\mathcal{H}_{m}} \mathcal{N}(i)
\vspace{-2pt}
\end{equation}
and $\mathcal{H}_{m}$ is the set containing all of $v$'s neighbors that forwarded a copy of packet $p_{m}$.

An important feature of a coding process is to be able to find coding opportunities that involve more than two native packets, i.e. increase what is known as the \emph{coding depth}. This feature is critical because it allows for further reduction of transmissions, thus improving energy efficiency. To illustrate that it is possible to use the proposed method to find coding opportunities involving multiple packets let us extend the example in Fig.~\ref{coding_opportunities_3_prev}. Assume now that another native packet $p_{3}$ is available at node $v$ and that we wish to check whether we can include it in the original coding $p_1$$\oplus$$p_2$, i.e. create the encoded packet $p_1$$\oplus$$p_2$$\oplus$$p_3$. 
Recall that the prerequisite is that every node in $\mathcal{N}(v)$ should know at least two of the thee packets. Observe that nodes in the set $\mathcal{S}=\mathcal{N}(w)\cap(\mathcal{N}(u)\cup\mathcal{N}(z))=\mathcal{Z}_{1}\cap\mathcal{Z}_{2}$ have received both $p_{1}$ and $p_{2}$ therefore they fulfill the prerequisite. On the other hand, nodes in $\mathcal{N}(v)-\mathcal{S}$ (gray area in Fig.~\ref{coding_opportunities_3_prev}) do not know both packets but have received either $p_{1}$ or $p_{2}$ (otherwise the coding of $p_{1}$ and $p_{2}$ could not be possible). Consequently, these nodes should known about $p_{3}$ in order for the triple coding to be possible. In other words, the set $\mathcal{R}\!\!=\!\!\mathcal{N}(v)-\mathcal{S}-\mathcal{Z}_{3}$ should be an empty set. The process can be repeated recursively to include more native packets in the encoding. In general, in order to include a native packet $p_{n}$ in an encoding that already contains packets $p_{1},p_{2},\ldots,p_{m}$ node $v$ should check whether
\vspace{-6pt}
\begin{equation}
\mathcal{R}=\mathcal{N}(v)\!-\!\mathcal{S}-\mathcal{Z}_{n} = \mathcal{N}(v)\!-\!\bigcap\limits_{j=1}^{m}\mathcal{Z}_{j}-\mathcal{Z}_{n}
\vspace{-6pt}
\end{equation}
is an empty set. Fig.~\ref{coding_pseudocode} presents the pseudocode of NOB-CR's coding procedure.

As we mentioned previously, our method renders the use of a neighbor reception table obsolete which significantly alleviates the related costs. Instead, we mostly rely on information already available through the underlying broadcast algorithm, i.e. neighborhood information. The only additional information that the method requires for a packet $p$ is the set of previous hops $\mathcal{H}_{p}$, i.e. the nodes that forwarded a copy of $p$ to $v$. This information can be used to estimate the set $\mathcal{Z}_{p}$ (procedure $GetReceiversOf(p)$ in the pseudocode) which consists of the nodes that have received $p$. This can be done using neighborhood information, i.e. $\mathcal{Z}_{p}=\bigcup_{\forall i \in \mathcal{H}_{p}} \mathcal{N}(i)$. Bear in mind that neighborhood information is updated on a periodic basis. Furthermore, each packet remains to the output queue for a limited period of time which is smaller than typical values for a neighborhood update interval. Therefore estimating the receivers of $p$ by using $\mathcal{Z}_{p}$ is as accurate as the neighborhood information. However, note that due to periodic updating and mobility, $\mathcal{Z}_{p}$ is an approximation of the nodes that actually received $p$. In Section~\ref{sec:Evaluation} we evaluate the impact of mobility on our coding approach and show that it is negligible even in networks of high node mobility. In addition, we examine in detail the pros and cons of using the proposed coding approach over the traditional one that utilizes a reception table.
\begin{figure}[!t]
	\begin{center}
		\line(1,0){250}
	\end{center}
	{
		\footnotesize
		\vspace{-14pt}
		$\;$\textbf{DetectCodingOpportunities(packet $p$, output Queue $\mathcal{Q}$)}
		\vspace{-17pt}
	}
	\begin{center}
		\line(1,0){250}
	\end{center}
	\vspace{-8pt}
	\begin{algorithmic}[1]
		\footnotesize
		
		\STATE $\mathcal{Z}_{p}=GetReceiversOf(p)$
		\STATE $\mathcal{S}=\mathcal{Z}_{p}$
		\STATE $\mathcal{C}=\mathcal{N}(v)-\mathcal{S}$
		\STATE $e=p$
		\FOR{each native packet \textbf{q} $\in \mathcal{Q}$}
		\STATE $\mathcal{Z}_{q}=GetReceiversOf(q)$
		\STATE $\mathcal{R}=\mathcal{C}-\mathcal{Z}_{q}$
		\IF{($\mathcal{R}=\emptyset$)}
		\STATE $\mathcal{S}=\mathcal{S}\cap\mathcal{Z}_{q}$
		\STATE $\mathcal{C}=\mathcal{N}(v)-\mathcal{S}$
		\STATE $e=e\oplus q$
		\ENDIF
		\ENDFOR
		\RETURN $e$
		\normalsize
	\end{algorithmic}
	\vspace{-20pt}
	\begin{center}
		\line(1,0){250}
	\end{center}
	\vspace{-15pt}
	\caption{Pseudocode for detecting coding opportunities at node $v$.}
	\label{coding_pseudocode}
	\vspace{-12pt}
\end{figure}

Clearly, the advantage of our method is the limited cost for storing and updating $\mathcal{H}_{p}$. To explain this note that the basic data item for representing either a neighbor reception table or $\mathcal{H}_{p}$ is the id of a node. Now observe that for each packet $p$ received by a node $v$ the neighbor reception table may store up to $|N(v)|$ items in contrast to the $|\mathcal{H}_{p}|$ items stored in our approach. By definition, in a broadcast algorithm $|\mathcal{H}_{p}|\ll|N(v)|$. The total storage gain is $n_{p}\times(|N(v)|-|\mathcal{H}_{p}|)$, where $n_{p}$ is the average number of packets for which reception information is stored at any given time. In Section~\ref{sec:Evaluation} we show that in our simulation set-up the storage requirement may be reduced by three orders of magnitude. Another benefit stemming from the limited storage requirement is the positive impact on the processing cost for updating this information (e.g. locating the appropriate $\mathcal{H}_{p}$ and adding a node id). Last but not least, the proposed algorithm for detecting coding opportunities is entirely based on the manipulation of sets. Therefore it is possible to represent all sets using bitmaps and implement our algorithm as a sequence of fast bitmap operations.

\subsection{Exploiting RAD to enhance the pruning process}\label{sec:Pruning}

The pruning efficiency of the underlying CDS-based algorithm is equally important to the coding process for minimizing transmissions. In general, algorithms that use the source-based CDS approach are considered to be the most efficient. In fact, NOB-CR builds on top of such an algorithm, i.e. PDP. Recall that in PDP a node $v$ elects forwarders from $\mathcal{C}(v)\!\!=\!\!\mathcal{N}(v)\!-\!\mathcal{N}(u)$ in order to deliver a packet $p$ to all nodes in $\mathcal{U}(v)=\mathcal{N}(\mathcal{N}(v))-\mathcal{N}(v)-\mathcal{N}(u)-\mathcal{N}(\mathcal{N}(u)\cap \mathcal{N}(v))$, where $u$ is the previous hop node that forwarded $p$ to $v$. We take advantage of information already provided by the RAD mechanism to further enhance the pruning efficiency of PDP. More specifically, we make the observation that, when electing the forwarders, it is possible to take into account not only the previous hop node of $p$ but all other nodes that relayed a duplicate of $p$ while it was buffered due to the RAD mechanism. As a result, the set of nodes to be covered can be further reduced to:
\begin{equation}\label{eq:NOB-CR_FW}
	\mathcal{U}(v)=\mathcal{N}(\mathcal{N}(v))-\mathcal{N}(v)-\bigcup\limits_{i\in\mathcal{H}_{p}}\mathcal{N}(i)-\bigcup\limits_{i\in\mathcal{H}_{p}}\mathcal{N}(\mathcal{N}(i)\cap \mathcal{N}(v))
\end{equation}
where the set $\mathcal{H}_{p}$ is the set consisting of all previous hop nodes of $p$. The same idea can be applied to any source-based CDS algorithm that uses the same problem modeling such as DP, TDP and their derivatives. It is clear that reducing $\mathcal{U}(v)$ increases the probability of selecting fewer forwarding nodes from the candidate set $\mathcal{C}(v)$. Also note that the proposed method comes at no additional cost since the set $\mathcal{H}_{p}$ is used for detecting coding opportunities.

\subsection{Reducing delay through coded redundancy}\label{sec:coded_redundancy}

\begin{figure}[!b]
	\centering
	\includegraphics[]{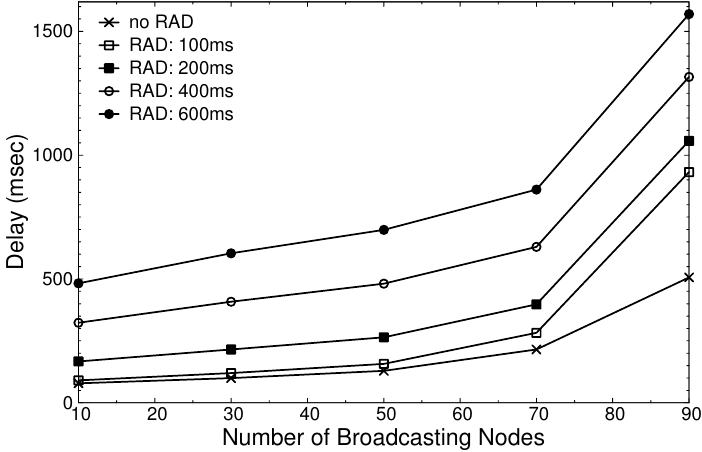}
	\caption{End-to-end delay performance of PDP MC/U under varying traffic load using different RAD intervals.}
	\label{PDP-MC_U-Delay}
\end{figure}
The performance of XOR-based broadcast schemes heavily depends on the random assessment delay (RAD) applied before relaying a packet. In particular, high RAD values increase the coding opportunities and maximize the coding gain. This behavior was confirmed by the experimental results in Section~\ref{sec:motivation} (Fig.~\ref{motivation_traffic_load_plots} and Fig.~\ref{motivation_scalability_plots}) where increasing the RAD value from $200$ to $400$ $ms$ substantially enhances CodeB's performance in terms of delivery ratio and energy efficiency. However, imposing a RAD to a packet in every node has a major impact on the end-to-end delay. Although RAD is usually short it results to an aggregation of a considerable end-to-end delay. To highlight this effect, we replayed the first experiment in Section~\ref{sec:motivation} and recorded the end-to-end-delay for different values of RAD (Fig.~\ref{PDP-MC_U-Delay}). To avoid any interference caused by the performance degradation of the M/U criterion under high load we used the PDP algorithm with the MC/U criterion which can sustain performance in such conditions (Fig.~\ref{PDP_MCU_ttl_traffic_load_plots}, Section~\ref{sec:MCU}). 

The results confirm our observation and expose the paramount importance of reducing the end-to-end delay when employing RAD. One way towards this direction is to increase the packet redundancy across the network. The rationale is that \textit{using more duplicates per packet increases the probability of delivering a copy of the packet through a faster path}. However, the practice of increasing redundancy should be exercised with caution because it usually results in extra transmissions. This impacts the energy efficiency of the broadcast process as well as its delivery efficiency through the increase of collisions.

We propose a cost-free method for introducing packet redundancy across the network. This method, called \emph{Coded Redundancy (CR)}, targets at increasing the duplicates of a packet and it is cost-free in the sense that it does not produce new transmissions. To accomplish that, CR introduces a new packet type called \emph{gratis}. Gratis packets are non-coded packets for which the receiving node has not been selected for relaying them. Instead of dropping them, our method examines if these packets could be forwarded as part of already encoded packets, i.e. without cost. Packets already delivered to all neighbors of a node $v$ do not qualify for marked as gratis because no delay improvement is feasible. Summarizing, node $v$ can mark gratis packets using the following criterion:
\begin{definition}[Gratis Marking Criterion]
	A native packet $p$ is marked as gratis by a node $v$ if there is at least one neighbor of $v$ that has not received $p$ and $v$ is not a forwarder of $p$.
\end{definition}
\noindent Note that it is possible for $v$ to estimate whether a packet has not been received by all of its neighbors by utilizing a similar methodology as the one used for detecting coding opportunities (Section~\ref{sec:coding_opp}).

Many aspects of packet handling are the same for gratis and native packets. More specifically, in order to avoid loops, gratis packets are considered for forwarding only if they conform to the termination criterion. Accepted gratis packets are also temporarily buffered using the RAD technique. Then, if a coding opportunity is detected one or more gratis packets are relayed as part of an already encoded packet. Nonetheless, there are also differences in handling gratis and native packets. The first is that gratis packets are dropped if their buffering time expires without finding a coding opportunity. This is because they are meant to be forwarded only as part of an encoded packet. For the same reason gratis packets do not participate in the forwarding process, i.e. a node does not determine a set of forwarders for a gratis packet. Instead, as soon as a packet is marked as gratis it is always treated as gratis in subsequent hops. This approach guarantees that gratis packets are forwarded without any additional cost.  
Finally, each node applies the following rule upon reception of a gratis packet:
\begin{definition}[Gratis Receiving Rule]
	The arrival of a gratis packet never triggers any modification to the termination criterion structures.
\end{definition}

\begin{figure}[!t]
	\centering
	{

		\begin{tikzpicture}[->,>=stealth',shorten >=1pt,auto,node distance=3cm,
		thick,main node/.style={circle,fill=beaublue,draw,font=\sffamily\Large\bfseries}]
		
		\node[main node,fill=lightgray] (1) at (1,0) {$u$};
		\node[main node,fill=white] (2) at (-1.7,0) {$v$};
		\node[main node,fill=lightgray] (3) at (1,2) {$z$};
		\node[main node,fill=lightgray] (11) at (-1.7,2) {$s$};
		\node[main node,fill=lightgray] (4) at (3.7,0) {$x$};
		\node[fill=white] (5) at (4.9,0) {\dots};
		\node[main node,fill=white] (6) at (6.25,0) {$a$};
		\node[fill=white, rotate=32] (7) at (4.7,0.55) {\dots};
		\node[main node,fill=white] (8) at (5.6,1.1) {$e$};
		\node[fill=white, rotate=61] (9) at (4.3,1) {\dots};
		\node[main node,fill=white] (10) at (4.8,1.9) {$n$};

		\node[] (t0a) at (-1.45,1.4) {\textbf{$t_0$:}};
		\fill [draw,thin,fill=white,rounded corners=2pt] (-1.2,1.2) rectangle (-0.8,1.6);
		\node[] (p1a) at (-0.97,1.4) {$p_1$};
		
		\node[] (t0b) at (-1.2,2.3) {\textbf{$t_0$:}};
		\fill [draw,thin,fill=white,rounded corners=2pt] (-1.0,2.1) rectangle (-0.6,2.5);
		\node[] (p1b) at (-0.77,2.3) {$p_1$};
		
		\node[] (t1) at (-1.2,0.3) {\textbf{$t_1$:}};
		\fill [draw,thin,fill=white,rounded corners=2pt] (-1.0,0.1) rectangle (0.5,0.5);
		\node[] (p1c) at (-0.25,0.3) {\textcolor{gray}{$p_1$}$\oplus$$p_2$$\oplus$$p_3$};
		
		\node[] (t2) at (1.5,0.3) {\textbf{$t_2$:}};
		\fill [draw,thin,fill=white,rounded corners=2pt] (1.7,0.1) rectangle (3.2,0.5);
		\node[] (p1d) at (2.45,0.3) {\textcolor{gray}{$p_1$}$\oplus$$p_4$$\oplus$$p_5$};
		
		\node[] (t4) at (1.25,1.4) {\textbf{$t_4$:}};
		\fill [draw,thin,fill=white,rounded corners=2pt] (1.5,1.2) rectangle (1.9,1.6);
		\node[] (p1e) at (1.7,1.4) {$p_1$};
		
		\node[] (t5) at (0.6,-0.6) {\textbf{$t_5$:}};
		\fill [draw,thin,fill=white,rounded corners=2pt] (0.8,-0.8) rectangle (1.2,-0.4);
		\node[] (p1f) at (1,-0.6) {$p_1$};
		\node[] (reason1) at (0.9,-1) {\footnotesize (due to termination criterion)};
		\fill [draw,-] (1,-0.6) node[cross=0.2cm,rotate=0,gray!80]{};
		
		\node[] (t3) at (3.3,-0.6) {\textbf{$t_3$:}};
		\fill [draw,thin,fill=white,rounded corners=2pt] (3.5,-0.8) rectangle (3.9,-0.4);
		\node[] (p1g) at (3.7,-0.6) {\textcolor{gray}{$p_1$}};
		\node[] (reason2) at (3.6,-1) {\footnotesize (no coding)};
		\fill [draw,-] (3.7,-0.6) node[cross=0.2cm,rotate=0,gray!80]{};
		
		\path[every node/.style={font=\sffamily\small}]
		(2) edge node [above] {} (1)
		(3) edge node [above] {} (1)
		(1) edge node [above] {} (4)
		(4) edge node [above] {} (5)
		(5) edge node [above] {} (6)
		(4) edge node [above] {} (7)
		(7) edge node [above] {} (8)
		(4) edge node [above] {} (9)
		(9) edge node [above] {} (10)
		(11) edge node [above] {} (2)
		(11) edge node [above] {} (3);
		
		\end{tikzpicture}
	}	
	\caption{Example of failing to deliver packet $p_1$ to all network nodes when the ``Gratis Receiving Rule'' is not implemented.}
	\label{example_gratis_failure}
	\vspace{-12pt}
\end{figure}
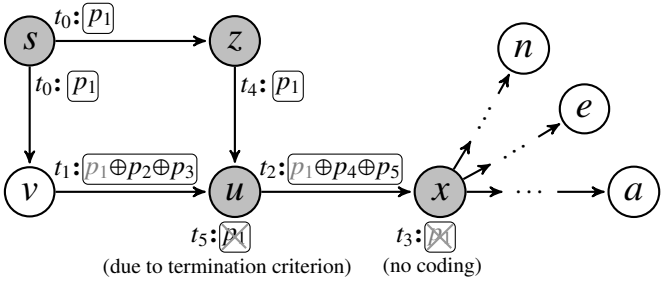
\noindent This rule is directly associated with the termination criterion and indirectly affects the other packet duplicates that coexist in the network. If not applied, it could lead to situations where packets prematurely terminate their propagation in the network destroying the protocol's delivery efficiency. We further investigate the importance of this rule through an example. In Fig.~\ref{example_gratis_failure}, we monitor a packet $p_1$ that is propagated through a part of a network. Nodes in gray are selected as forwarders for $p_1$, while all other nodes act as passive receivers. Initially, at time $t_0$, nodes $v$ and $z$ receive a duplicate of $p_1$ from the source node $s$. Node $v$ handles $p_1$ as gratis since it is not elected as a forwarder while for the opposite reason node $z$ treats $p_1$ as native. At some point in time ($t_1$), node $v$ detects a coding opportunity between the gratis $p_1$ and the already encoded packet $p_2$$\oplus$$p_3$. As a result, it transmits the encoded packet $p_1$$\oplus$$p_2$$\oplus$$p_3$. Node $u$ receives the encoded packet and successfully decodes it (assuming $p_2$ and $p_3$ are already known). Node $u$ also handles the copy of $p_1$ as gratis. At time $t_2$ a coding opportunity for gratis packet $p_1$ at node $u$ results in the transmission of a new encoded packet, i.e. $p_1$$\oplus$$p_4$$\oplus$$p_5$, where $p_4$ and $p_5$ are ordinary native packets previously received at node $u$. Node $x$ receives the encoded packet and decodes it (assuming $p_4$ and $p_5$ are already known). Likewise, $x$ handles $p_1$ as gratis searching for a proper coding opportunity to relay it. Assuming that $p_1$'s buffering time expires with no coding opportunities, $x$ drops $p_1$ (time $t_3$), terminating its dissemination to the rest of the network. At this point, the only way to deliver $p_1$ to nodes $n$, $e$ and $a$ is through the duplicate of $p_1$ that node $z$ holds. At time $t_4$, the duplicate of $p_1$ is transmitted as a native packet because $z$ is an elected forwarder for $p_1$. Node $u$, which is also an elected forwarder, successfully receives $p_1$ and has the opportunity to further relay it. However, its decision depends on the previous reception of the gratis copy of $p_1$ at time $t_1$. In particular, $u$ employs the MC/U termination criterion which allows a node to relay only packets seen for the first time. In case that $u$ has recorded the former arrival of $p_1$ at $t_1$ no forwarding is allowed due to the termination criterion. Consequently, the newly fetched duplicate of $p_1$ is dropped (time $t_5$). On the other hand, implementing the ``Gratis Receiving Rule'' resolves the situation. According to the rule, the former arrival of $p_1$ as a gratis packet is never registered by node $u$. As a result, the native copy of $p_1$ is forwarded to node $x$ (time $t_5$). From that point, node $x$ successfully propagates $p_1$ across all parts of the network.

The example in Fig.~\ref{example_gratis_failure} also illustrates the benefits of using the coding redundancy technique. Packet $p_1$ reaches nodes $u$ and $x$ much faster when coding redundancy is utilized. More specifically, nodes $u$ and $x$ receive $p_1$ at time instances $t_1$ and $t_2$, respectively. On the contrary, without coded redundancy, $p_1$ reaches nodes $u$ and $x$ at $t_4$ and $t_5$, respectively.

The key functionality of the coding operation is the detection of coding opportunities. Extending this functionality to support gratis packets is not straightforward. This is because there are cases where gratis packets could destroy coding opportunities involving native ones, thus impairing the protocol's energy efficiency. Let us examine this problem through an example. Suppose that an intermediate node $v$ is elected as forwarder for two native packets, i.e. $p_{1}$ and $p_{2}$, and that these packets can be mixed together (forming the encoded packet $p_{1}$$\oplus$$p_{2}$) and forwarded with a single transmission. Things get complicated when a gratis packet $p_{3}$ is also present at $v$. Assuming that $p_{3}$ can be combined only with $p_{1}$ and that $v$ chooses to combine $p_{1}$ with $p_{3}$ instead of $p_{2}$ then two transmissions are required, i.e. one for $p_{1}$$\oplus$$p_{3}$ and one for $p_{2}$. To evade this problem, we introduce the following rule that every node applies when searching for coding opportunities.

\begin{definition}[Gratis Coding Rule]
	Gratis packets participate in the coding process with lower priority than native packets.
\end{definition}

Furthermore, in case that no coding opportunity is found among native packets then coding detection terminates without examining gratis packets. This policy in conjunction with the above rule prevents a gratis packet from directly being combined with existing native ones, leaving them available for possible future encodings. The only exception of combining a gratis packet with a native one is when the buffering time of the latter expires. In that case there is no possibility for the native packet to participate in any future coding opportunity. As a result, encoding the native packet with the gratis one does not destroy future coding opportunities.

\section{Evaluation}\label{sec:Evaluation}

\begin{table}[!b]
	\vspace{-12pt}
	\caption{Simulation parameters}
	\centering
	\begin{tabular}{|l|l|}
		\hline
		Simulation Time & $300$ s\\
		Number of Trials & $20$\\
		Confidence Level & $95$\% \\
		\hline \hline
		Transmission Range ($R$) & $250$ m\\
		Bandwidth & $2$ Mb/s\\
		Number of Nodes ($N$) & $60$ - $250$\\
		Avg. Neigh. Size & $15$, $30$\\
		Node Speed & $0$ - $20$ m/s \\
		Broadcast Sessions ($S$) & $10$ - $90$\\
		Broadcast Rate ($ \lambda $)& $0.1$ - $8$ pkts/s\\
		Packet Size & $256$ Bytes\\
		Hello Interval ($ t_{H} $) & $1$ s\\
		Random Assessment Delay ($ t_{RAD} $) & $100$ - $600$ ms\\
		\hline 

	\end{tabular}
	\label{SimParameters}
\end{table}

\begin{figure*}[!t]
	\centering
	\subfloat[]{\includegraphics[width=0.25\linewidth]{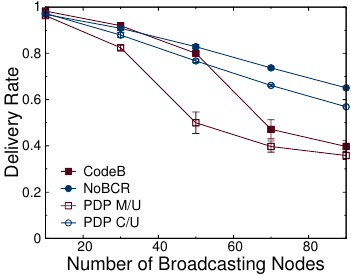}\label{sparse_DR}}
	\subfloat[]{\includegraphics[width=0.25\linewidth]{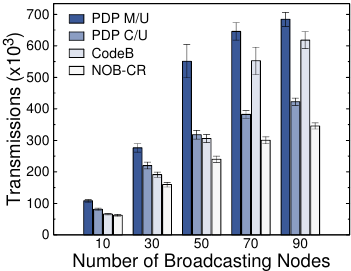}\label{sparse_TR}}
	\subfloat[]{\includegraphics[width=0.25\linewidth]{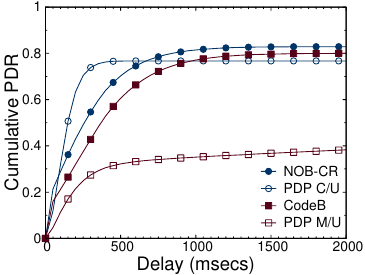}\label{sparse_CDF_50}}
	\subfloat[]{\includegraphics[width=0.25\linewidth]{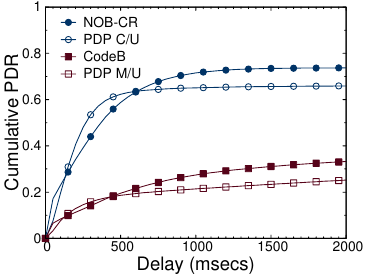}\label{sparse_CDF_70}}%
	\caption{Performance for different levels of offered load in the ``sparse" topology ($N\!\!=\!\!100$, max speed:1 m/s, $t_{RAD}\!\!=\!\!400$ms): (a) Delivery rate vs broadcasting sources (b) Avg. number of transmissions vs broadcasting sources (c) Cumulative PDR vs end-to-end delay ($S\!\!=\!\!50$ sources) (d) Cumulative PDR vs end-to-end delay ($S\!\!=\!\!70$ sources).}
	\vspace{-12pt}
	\label{sparse_topology}
\end{figure*}

To evaluate the performance of NOB-CR, we compare it with two algorithms. The first one is CodeB~\cite{CodeB} which is the most representative of XOR coding-based broadcast algorithms. The second algorithm is PDP~\cite{PDP} which, despite the fact that does not use any type of coding, is well-known for its energy efficiency. We use two variants of PDP, namely PDP M/U and PDP C/U, in order to examine how the termination criterion affects the overall performance.\\
\textit{Set up and methodology}: 
All investigated algorithms are implemented in the ns2 simulator~\cite{ns2}, using the CMU extension. We present the  average values over $20$ independent simulation runs, each with a duration of $300$ s. The confidence level, for all reported confidence intervals, is $95\%$.\\
\textit{Network model}:
The default number of nodes is $100$, the propagation model is the TwoRay ground with a transmission range of $250$m and the nominal bit rate is $2$Mbps. The nodes move in a square area according to the Random Waypoint (RW) model~\cite{rwp-dist}. To avoid transient artifacts in nodes' movement, we use the perfect simulation algorithm~\cite{prf-sim}. We examine two network topologies; ``dense" and ``sparse". Similar to \cite{CodeB}, in the ``dense" topology, the average neighborhood size is 30 while in the ``sparse" topology it is 15. Note that we could not use a lower density in the ``sparse" scenario since in such a case frequent partitions occur. Simulations confirmed that in the ``sparse" scenario there exist many nodes (those moving near the boundaries) that experience very low connectivity. All algorithms collect neighborhood information by periodically exchanging hello messages with an interval ($t_{H}$) of 1 s.\\
\textit{Network traffic}: 
Traffic is generated by broadcast sessions, each stemming from a different source node and starting at a random time. Although we use a variable number of sources, each one producing packets at a constant rate of $\lambda\!=\!1$pkt/s, the default value is $50$. The size of each message is set to $256$ Bytes.\\
\textit{Coding parameters}: 
All coding schemes under evaluation use the RAD technique to maximize the probability of coding opportunities. According to RAD, each node delays every packet it receives for a random delay in $[0,t_{RAD}]$. The default $t_{RAD}$ value in our simulations is $400$ ms. Due to storage limitations, all coding based algorithms buffer incoming packets for a limited time interval ($B_{T}$) in order to enable  decoding. This time interval is highly correlated with the time period for which information inside the neighbor reception table is available ($R_{T}$). We set both time intervals to $5$ s for the CodeB algorithm in order to increase the benefits of network coding and be realistic at the same time. On the contrary, we set only the $B_{T}$ interval for NOB-CR algorithm since it operates without a neighbor reception table. After extensive experimentation, we found that a $B_{T}$ value equal to $2$ s is sufficient for NOB-CR's encoding/decoding operation. The small $B_{T}$ value utilized by NOB-CR is preferable because it allows for efficient management of the limited storage space in the network nodes.

Table~\ref{SimParameters} summarizes the simulation parameters.

\begin{table}[!b]
	\vspace{-12pt}
	\caption{Reduction ($\%$) of transmissions for NOB-CR compared to other algorithms.}
	\centering
	\begin{tabular}{|c|l|l|l|}
		\hline
		\# sources & PDP M/U & PDP C/U & CodeB\\
		\hline \hline
		10 & 44.7\% & 27.0\% & 7.7\%\\
		30 & 44.1\% & 29.8\% & 17.9\%\\
		50 & 57.8\% & 26.9\% & 22.7\%\\
		70 & 55.1\% & 24.2\% & 45.5\%\\
		90 & 51.3\% & 21.1\% & 45.3\%\\
		\hline 
	\end{tabular}
	\label{sparse_energy_efficiency}
\end{table}
Fig.~\ref{sparse_topology} illustrates the performance of all investigated algorithms in the sparse topology and under different levels of offered load (variable number of broadcasting sources). As discussed in section~\ref{sec:motivation}, our experimental results reveal the ineffectiveness of the M/U criterion that induces the performance breakdown of the PDP M/U and CodeB schemes. More specifically, as the load increases the algorithms that utilize the M/U criterion lose their pruning efficiency producing a large number of transmissions (Fig.~\ref{sparse_TR}). PDP M/U fails to prune transmissions when the number of source nodes exceeds $30$. On the other hand, network coding enables CodeB to maintain its pruning efficiency when traffic is produced by up to $50$ sources (half of network nodes). However, as the congestion level increases, the excessive number of forwarding decisions taken by both algorithms induce transmission failures due to packet collisions. As a result, their delivery performance deteriorates (Fig.~\ref{sparse_DR}). NOB-CR outperforms all algorithms both in terms of delivery efficiency and number of transmissions. Even in the extreme case of $90$ broadcasting sources, which is close to the all-to-all communication paradigm, NOB-CR delivers $\sim$$66\%$ of the traffic while the M/U based schemes reach less than $40\%$ of the network nodes (Fig.~\ref{sparse_DR}). This justifies our approach to combine XOR network coding with an termination criterion other than M/U. At the same time, NOB-CR is exceptionally energy efficient. Table~\ref{sparse_energy_efficiency} presents NOB-CR's energy gains that derive from reducing the total number of transmissions. Against PDP variants, NOB-CR reduces the total number of transmissions by $21\%$ in the worst case. Compared to CodeB, the energy gains become noticeable when the broadcasting sources are more than $30$ ($18\%$ to $45\%$).

Fig.~\ref{sparse_CDF_50} and~\ref{sparse_CDF_70} depict the cumulative packet delivery ratio (PDR) versus the end-to-end delay, i.e. the fraction of packets delivered within a delay limit, when the broadcasting sources are $50$ and $70$, respectively. We choose this presentation style in order to capture both the delivery efficiency and the timeliness of each algorithm. Again, the results provide a confirmation of the ineffectiveness of the M/U criterion. Even when there are only $50$ sources in the network (Fig.~\ref{sparse_CDF_50}) PDP M/U collapses while XOR coding allows CodeB to achieve a competitive performance. However, when the offered load increases (Fig.~\ref{sparse_CDF_70}), both schemes that use the M/U criterion deliver less packets with higher delay due to the increased number of transmission failures. NOB-CR outperforms both schemes not only in delivering more packets, but in delivering them faster. The reason is twofold; the utilization of the MC/U criterion and the coded redundancy mechanism. Interestingly, NOB-CR's delay profile is comparable to that of PDP C/U (especially in the case of 70 sources) despite the fact that simple PDP schemes operate without the RAD mechanism that significantly increases the end-to-end delay. Moreover, we found that under high load ($>\!\!70$ sources) NOB-CR is at least as fast as PDP C/U. This is due to NOB-CR's pruning process that efficiently decreases transmission failures allowing for timely packet delivery.

We also experimented on increasing the packet generation rate $\lambda$ while keeping the number of sources constant, e.g. $S\!\!=\!\!10$ sources. In this way it is possible to change the offered load but limit the coding capability of algorithms. This is because coding opportunities heavily depend on the number of packet flows, i.e. the number of sources. 
The obtained results are qualitatively similar to the previous experiment therefore we do not include the corresponding performance plots. In summary, for a low $\lambda$, the energy gains for all coding enabled schemes are limited. This is because the number of coding opportunities is rather small since fewer packets coincide in the network. As the offered load increases, the benefits of network coding become more evident. However, after the breaking point of $\lambda\!\!=\!\!5$ packets per second the delivery performance of CodeB deteriorates as a result of the increased levels of congestion. On the other hand, NOB-CR exhibits a remarkable resilience to congestion achieving the best performance in terms of delivery ratio and energy efficiency.

\begin{figure}[!t]
	\centering
	{
		\subfloat[Case I][]
		{
			\includegraphics[width=0.47\linewidth]{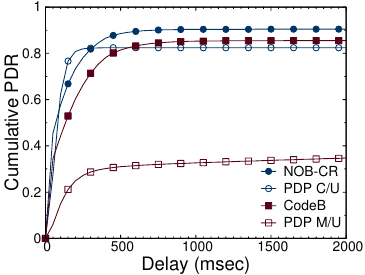}
			\label{dense_CDF_50}
		}
		\subfloat[Case II][]
		{
			\includegraphics[width=0.47\linewidth]{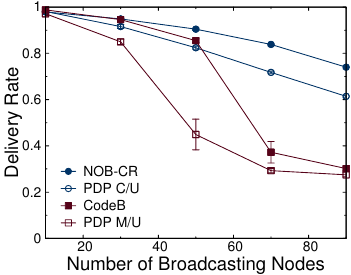}
			\label{dense_DR}
		}
		\hfil
		\subfloat[Case III][]
		{
			\includegraphics[width=0.47\linewidth]{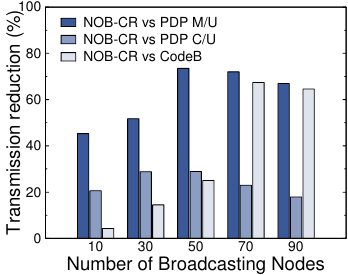}
			\label{dense_energy_gains}
		}
	}
	\caption{Performance for different levels of offered load in the ``dense" topology ($N\!\!=\!\!100$, max speed:1 m/s, $t_{RAD}\!\!=\!\!400$ms): (a) Cumulative PDR vs end-to-end delay ($S\!\!=\!\!50$ sources) (b) Delivery rate vs sources (c) Transmission reduction vs sources.}
	\vspace{-12pt}
	\label{dense_topology}
\end{figure}

Next, we used a variable number of sources to test all schemes under different levels of offered load in the dense topology (Fig.~\ref{dense_topology}). As expected, when the offered load is low, the delivery efficiency of all algorithms improves compared to the sparse topology (Fig.~\ref{dense_CDF_50}). This is because the higher number of neighbors results in increased packet redundancy, making delivery more probable. Furthermore, the diameter of a denser network is smaller.
Consequently, as depicted in Fig.~\ref{dense_CDF_50}, all schemes deliver packets faster than in the sparse case (Fig.~\ref{sparse_CDF_50}). However, despite its positive effects, there is also a downside of the increased neighborhood size; under high load the probability of transmission failures due to collisions is higher. This is because more packet duplicates are created, resulting in congestion. As a result, schemes that do not efficiently prune transmissions collapse (Fig.~\ref{dense_DR}) as the offered load increases. Interestingly, the performance degradation is more acute and develops more quickly (lower traffic levels) than in the case of sparse topology because congestion is more severe. Fig.~\ref{dense_energy_gains} illustrates the reduction of the total number of transmissions achieved by NOB-CR compared to all other schemes. As anticipated, the higher gains are witnessed when the offered load is high where NOB-CR prunes $60\%$ more transmissions than CodeB. At the same time, it delivers over $40\%$ more packets (Fig.~\ref{dense_DR}).

\begin{figure}[!b]
	\vspace{-12pt}
	\centering
	{
		\subfloat[Case I][]
		{
			\includegraphics[width=0.47\linewidth]{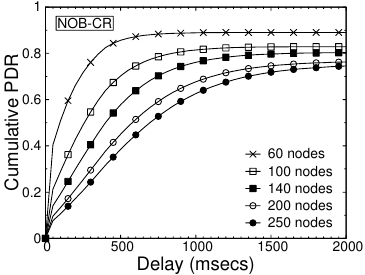}
			\label{scalability_CDF_NOB-CR}
		}
		\subfloat[Case II][]
		{
			\includegraphics[width=0.47\linewidth]{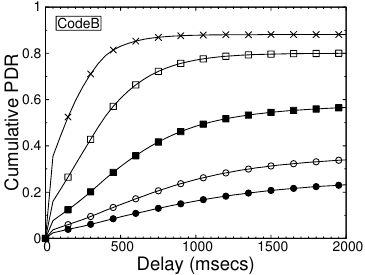}
			\label{scalability_CDF_CodeB}
		}
	}
	\vspace{-6pt}
	\caption{Performance when the network size scales up (``sparse" topology, $S\!\!=\!\!50$ sources, max speed:1 m/s, $t_{RAD}\!\!=\!\!400$ms): Cumulative PDR vs end-to-end delay for a) NOB-CR, and (b) CodeB.}
	\label{scalability_CDF}
\end{figure}
\begin{figure}[!t]
	\centering
	{
		\subfloat[Case I][]
		{
			\includegraphics[width=0.47\linewidth]{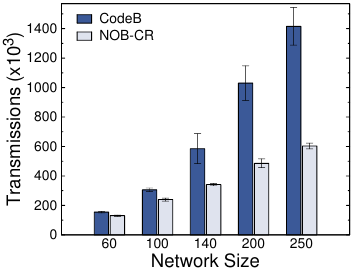}
			\label{scalability_TR}
		}
		\subfloat[Case II][]
		{
			\includegraphics[width=0.47\linewidth]{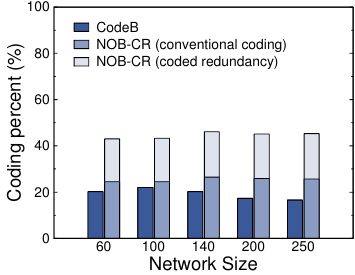}
			\label{scalability_Coding}
		}
	}
	\caption{Performance when the network size scales up (``sparse" topology, $S\!\!=\!\!50$ sources, max speed:1 m/s, $t_{RAD}\!\!=\!\!400$ms): (a) Avg. number of transmissions, and (b) Fraction of encoded packets vs network size.}
	\vspace{-12pt}
	\label{scalability_pruning}
\end{figure}

In the following we focus on the more challenging scenario of sparse networks. The next set of experiments assesses the performance of NOB-CR and CodeB when scaling the network up. Towards this direction, we conducted simulations with an increasing number of nodes. At the same time, we also expand the network area in which the nodes move in order to keep the average neighborhood size fixed. Fig.~\ref{scalability_CDF} illustrates the cumulative packet delivery ratio versus the end-to-end delay for various network sizes. NOB-CR and CodeB present a similar behavior when the network size is small, i.e. $60$ nodes. As the number of nodes increases, the performance of CodeB quickly deteriorates and finally collapses when the network size exceeds $140$ nodes (Fig.~\ref{scalability_CDF_CodeB}). On the other hand, NOB-CR exhibits a remarkable durability and its performance, in terms of both delivery ratio and end-to-end delay, degrades much more slower and smoother (Fig.~\ref{scalability_CDF_NOB-CR}). The witnessed performance degradation for both schemes is reasonable since in our experiment we fix the network density. As a result, the diameter of the network increases with the number of nodes and therefore it is more difficult to reach some destinations. Notwithstanding, NOB-CR is very efficient in reducing transmissions (Fig.~\ref{scalability_TR}) and therefore alleviates congestion. Thus, failures due to collisions are minimized and so is the impact of the increasing network diameter.
For example, in the case of $60$ nodes NOB-CR produces $\sim$$16\%$ less transmission than CodeB, while in case of $140$ nodes transmissions are reduced by $\sim$$42\%$. This increasing difference in the pruning efficiency of the two schemes not only justifies the higher deliver ratio of NOB-CR but is also in accordance with the increasing difference in the delivery efficiency of the two schemes. Overall, NOB-CR loses less than $15\%$ of its delivery efficiency in the most demanding scenario of $250$ nodes. On the other hand, in the same scenario CodeB loses more than $60\%$.

Going back to Fig.~\ref{scalability_TR}, it is worth pointing out that the pruning efficiency of NOB-CR is increasingly better compared to that of CodeB. As expected, the total number of transmissions increases because more forwarders are required for a bigger network with a larger diameter. However, NOB-CR manages to suppress this increase and therefore broaden its advantage over CodeB. The reason for this result is not only the better pruning operation of the MC/U termination criterion but also its more efficient coding operation. To illustrate this, we present in Fig.~\ref{scalability_Coding} the number of transmitted encoded packets as a percentage of the total (encoded and native) number of transmitted ones. Furthermore, in the case of NOB-CR we present two separate classes on encoded packets. The one consists of encoded packets containing at least one gratis packet (Coded Redundancy) while the other refers to typical encodings involving only native packets (conventional coding). Recall that besides the packets encoded with the conventional mechanism, those produced with the Coded Redundancy method may also reduce the number of transmissions. This is because the latter packets may also contain two or more native packets.
Clearly, NOB-CR not only consistently performs about twice as many encodings as CodeB does, but its coding operation is also stable. 
In contrast, the percentage of coded packets for CodeB decreases in bigger networks which implies that the coding operation is hampered by the underlying broadcast mechanism. Another important finding is that NOB-CR also performs a significant number of encodings involving gratis packets. Although this type of encodings is important for reducing end-to-end delay (as discussed in Section~\ref{sec:coded_redundancy}), it is also beneficial for enhancing delivery efficiency because gratis packets increase packet redundancy across the network without any additional cost. Therefore, the Coded Redundancy mechanism also contributes to NOB-CR's superior delivery efficiency witnessed in Fig.~\ref{scalability_CDF}.

\begin{figure}[!t]
	\centering
	{
		\subfloat[Case I][]
		{
			\includegraphics[width=0.47\linewidth]{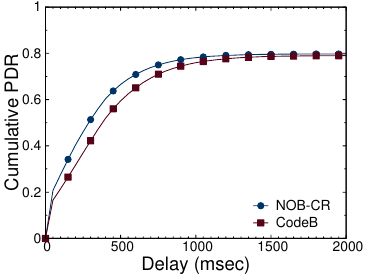}
			\label{mobility_CDF_2-10}
		}
		\subfloat[Case II][]
		{
			\includegraphics[width=0.47\linewidth]{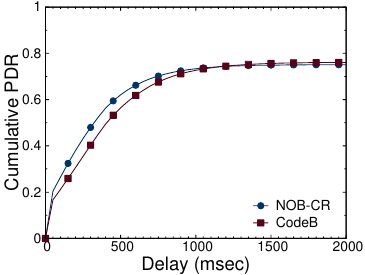}
			\label{mobility_CDF_10-20}
		}
	}
	\caption{Performance under different levels of mobility (``sparse" topology, $N\!\!=\!\!100$, $S\!\!=\!\!50$ sources, $t_{RAD}\!\!=\!\!400$ms): Cumulative PDR vs end-to-end delay for Node speed: (a) $2-10$ m/s (b) $10-20$ m/s.}
	\label{mobility_CDF}
\end{figure}

In the last experiment we assess the delivery efficiency under different levels of mobility (Fig.~\ref{mobility_CDF}). Clearly, increased mobility impacts the performance of both NOB-CR and CodeB. The reason is that both schemes rely on the PDP scheme. The latter uses neighborhood information for electing the optimal forwarders. This information becomes outdated more quickly when mobility increases. Note that, as discussed in Section~\ref{sec:coding_opp}, NOB-CR uses neighborhood information also for detecting coding opportunities while CodeB uses a neighbor reception table. Nonetheless, NOB-CR manages to outperform CodeB regardless of the mobility level and even in scenarios of very high mobility. More specifically, it exhibits faster packet delivery (Fig.~\ref{mobility_CDF}) using more than $\sim$$20\%$ less transmissions compared to CodeB (Table~\ref{mobility_energy_efficiency}). 
\begin{table}[!t]
	\caption{ NOB-CR's energy gains over CodeB under different mobility levels}
	\centering
	\begin{tabular}{|c|c|c|c|}
		\hline
		\textbf{Mobility} & $0-1$ m/s & $2-10$ m/s & $10-20$ m/s\\
		\hline \hline
		\textbf{Gain} & 21.45\% & 23.94\% & 27.83\%\\
		\hline 
	\end{tabular}
	\label{mobility_energy_efficiency}
\end{table}

Besides the performance evaluation of the complete NOB-CR algorithm it is interesting to shed some more light on the advantages and the limitations of the lightweight coding mechanism proposed in Section~\ref{sec:coding_opp}. In other words, we wish to investigate the storage and processing gains as well as the coding efficiency of the approach, i.e. the ability to find coding opportunities without those ending up in decoding failures, compared to the traditional approach that uses a neighbor reception table. To this end and in order to rule out any other interfering factor, we compare NOB-CR with a modified version of it that uses the typical neighbor reception table instead of the lightweight coding mechanism. 
\begin{figure}
	\centering
	\includegraphics[]{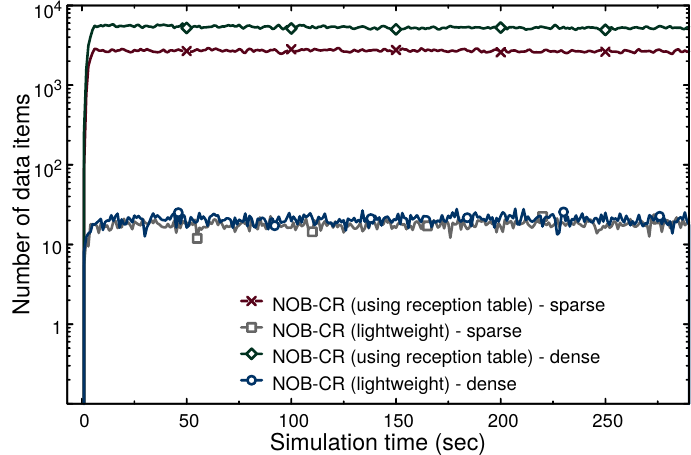}
	\vspace{-6pt}
	\caption{Average number of items stored by a node over simulation time when using neighbor reception table vs lightweight implementation for ``sparse" and ``dense" topologies (max speed: $1$ m/s, $N\!\!=\!\!100$, $S\!\!=\!\!50$, $t_{RAD}\!\!=\!\!400$ms).}
	\label{light_vs_normal_elements}
	\vspace{-6pt}
\end{figure}
Clearly the advantage of the lightweight coding mechanism is the reduced storage and processing requirement as discussed in Section~\ref{sec:coding_opp}. To quantify this advantage we monitored the storage requirement for the two coding schemes both in a ``sparse" and a ``dense" network (Fig.~\ref{light_vs_normal_elements}). We express the storage requirement in terms of data items, where a data item represents the memory required for storing the id of a node. We follow this approach in order to have a fair comparison that does not depend on the data representation method. Evidently, the storage demand for the lightweight approach is significantly smaller (up to three orders of magnitude) compared to the case of a neighbor reception table. As discussed in Section~\ref{sec:coding_opp}, this also has a positive impact on the required processing. An interesting and useful feature is that although the storage demand of the traditional coding approach increases (almost doubles) in a dense topology this is not true for the lightweight implementation. This is reasonable because in the latter case the storage demand depends on the rather stable number of received duplicates and not on the neighborhood size. 

\begin{figure}[!t]
	\centering
	{
		\subfloat[Case I][]
		{
			\includegraphics[width=0.47\linewidth]{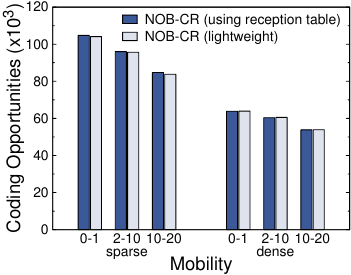}
			\label{light_vs_normal_mob_codopp}
		}
		\subfloat[Case II][]
		{
			\includegraphics[width=0.47\linewidth]{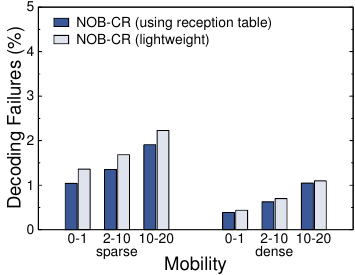}
			\label{light_vs_normal_mob_decodfail}
		}
	}
	\vspace{-6pt}
	\caption{Coding Efficiency of lightweight implementation vs neighbor reception table with respect to mobility in ``sparse" and ``dense" topologies ($N\!\!=\!\!100$, $S\!\!=\!\!50$, $t_{RAD}\!\!=\!\!400$ms): (a) coding opportunities (b) percentage of decoding failures.}
	\label{light_vs_normal_mob}
\end{figure}
Regarding the coding efficiency and in order to understand the differences between the two approaches recall that the neighbor reception table is populated upon the reception of a packet, i.e. it uses the neighborhood information at the time of packet reception (e.g. $t_{0}$). Instead, our method uses the neighborhood state at a later time $t_{1}>t_{0}$ (when a coding opportunity is present) as an estimation of the neighborhood at $t_{0}$. Clearly, this estimation becomes less accurate when mobility increases due to the increased invalidation rate of neighborhood information or when a packet waits a longer time for a coding opportunity (i.e. $t_{1}-t_{0}$ increases). We expect the lightweight coding operation to be challenged in the aforementioned conditions so we investigate the extent at which this happens. First we compare the two schemes under different levels of node mobility (Fig.~\ref{light_vs_normal_mob}). Furthermore, we examine both ``sparse" and ``dense" topologies that correspond to different neighborhood sizes. As expected, using a neighbor reception table is slightly better than the proposed lightweight method from a coding point of view, i.e. in terms of both detected coding opportunities (Fig.~\ref{light_vs_normal_mob_codopp}) and decoding failures (Fig.~\ref{light_vs_normal_mob_decodfail}). However, the difference is minor even in high levels of mobility. What is more important is that the slightly better coding operation of NOB-CR with reception table translates to a poor improvement in delivery ratio (Table~\ref{light_vs_traditional_dr}) which peaks at $\sim$$0.65\%$. This along with the advantages of the lightweight implementation justifies our approach to choose the latter over the traditional coding approach.
\begin{table}[!t]
	\caption{Delivery Ratio Reduction (\%) of NOB-CR compared to NOB-CR with neighbor reception table}
	\centering
	\begin{tabular}{|c|c|c||c|c|}
		\hline
		\textbf{Mobility} & & {\footnotesize DR Reduc. (\%)} & \textbf{RAD} & {\footnotesize DR Reduc. (\%)}\\
		\hline 
		0-1 m/s & \multirow{3}{*}{\rotatebox[origin=c]{90}{ ``dense'' }} & 0.1002 & 100 ms & 0.1342\\
		\hhline{-~---}
		2-10 m/s & & 0.0686 & 200 ms& 0.3426\\
		\hhline{-~---}
		10-20 m/s & & 0.0921 & 400 ms& 0.5302\\
		\hline
		0-1 m/s & \multirow{3}{*}{\rotatebox[origin=c]{90}{ ``sparse'' }} & 0.5302 & 600 ms& 0.8232\\
		\hhline{-~---}
		2-10 m/s & & 0.5677 &  & \\
		\hhline{-~---}
		10-20 m/s & & 0.6474 & &\\
		\hline 
	\end{tabular}
	\vspace{-6pt}
	\label{light_vs_traditional_dr}
\end{table}
\begin{figure}[!b]
	\vspace{-12pt}
	\centering
	{
		\subfloat[Case I][]
		{
			\includegraphics[width=0.47\linewidth]{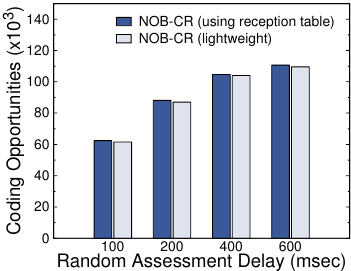}
			\label{light_vs_normal_rad_codopp}
		}
		\subfloat[Case II][]
		{
			\includegraphics[width=0.47\linewidth]{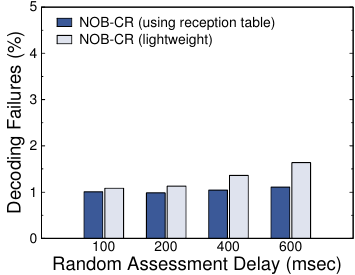}
			\label{light_vs_normal_rad_decodfail}
		}
	}
	\vspace{-6pt}
	\caption{Coding Efficiency of lightweight implementation vs neighbor reception table with respect to RAD (``sparse" topology, max speed: $1$ m/s, $N\!\!=\!\!100$, $S\!\!=\!\!50$): (a) coding opportunities (b) percentage of decoding failures.}
	\label{light_vs_normal_rad}
\end{figure}
Similar results are witnessed when we modify the RAD value, that is the maximum time that we allow a packet to wait for a coding opportunity (Fig.~\ref{light_vs_normal_rad}). As explained previously, in the lightweight approach late coding opportunities, i.e. those appearing significantly later than the reception of the involved packets, run an increased risk of resulting to a decoding failure. However, this does not significantly impact the performance (Table~\ref{light_vs_traditional_dr}). Even when RAD is as high as $600$ $ms$ the delivery ratio reduction is as low as $\sim$$0.82\%$ compared to the traditional case. Note that $600$ $ms$ is already a significantly high RAD value and results in increased end-to-end delay (Fig.~\ref{PDP-MC_U-Delay}). Using a higher value would further increase end-to-end delay thus destroying the broadcast process. Finally, we observed similar results when increasing the network size or the number of broadcasting sources.

\section{Related work}\label{sec:Related}
Several studies have explored the use of network coding for broadcasting in wireless ad hoc networks. Before looking into the proposed schemes we first briefly review the field of traditional (non-coded) broadcasting that has been extensively studied over the last years. The interested reader can refer to a set of surveys~\cite{survey-Williams,survey-Ruiz,survey-Stojmenovic-Wu,MPR-survey,survey-probabilistic} for more details.

\subsection{Traditional Broadcasting}

The challenge in broadcasting is to deliver a message to all network nodes while only a subset of the network nodes, called \emph{forwarders}, relay the message. Energy efficiency results from minimizing the set of forwarders. The simplest approach is to choose the set of forwarders probabilistically~\cite{survey-probabilistic}. However, more efficient approaches follow a deterministic approach by constructing a connected dominating set (CDS) of the network in a distributed fashion.
The nodes constituting the CDS are the potential forwarders while all other nodes just act as passive receivers. Then, extra rules are usually applied in order to select a subset of the CDS nodes as the forwarders.
	
Deterministic broadcast approaches can be classified into three broad categories. In the first, a CDS of the network is locally built using local topology information, i.e., 1-hop and 2-hop neighborhood. The computed CDS is used to forward broadcast packets throughout the network with packets stemming from different sources using the same CDS. Most algorithms in this category differentiate on the heuristics used for constructing the CDS~\cite{WuLi,DaiWu,it-cds,MPR-survey}.
Algorithms in the second category again locally build a CDS that is common to every network node but use additional dynamic rules based on broadcast state information to reduce the initial CDS. More specifically, dynamic rules usually exploit reception of packet duplicates to compute nodes that already received the packet and enhance the pruning process~\cite{SBA,DS-NES,DS-NES-supplementary,PBSM,ABSM,MPR,MPR2}. Other algorithms in this category focus on reliability either by introducing packet acknowledgements~\cite{ABSM,ABSM-theoretical} or by modifying the construction of the CDS~\cite{MPR-reliable}. Finally, the third category follows a different approach. Instead of building an initial CDS and then pruning it, the algorithms in this category construct a source-specific CDS on a hop-by-hop basis as packets are spread throughout the network~\cite{DP,PDP,Khabbazian}. The CDS in constructed using both local topology information, i.e, 1-hop and 2-hop neighbors, and broadcasting state information obtained through packet duplicates. Algorithms that focus on reliability also exist in this category~\cite{DCB}.

\subsection{Coding-based Broadcasting}

In the field of network-coded broadcast the proposed algorithms can be classified into: i) energy efficient~\cite{CodeB,NCDS_conf,NCDS_journal,kunz-iwcmc,OstovariXOR,Directional,fragouli_rlnc,widmer-extreme-net,mahmood_arlnccf_icc,RLDP}, and ii) delivery guarantee~\cite{DiSC,WangXOR,rahnavard2008CRBCast,vellambi2010FTS,widmer_rlnc-update,hou2008adapcode,cho_DRAGON,yang2011-RCODE-jrn,subramanian2012uflood,Chachulski-more,Koutsonikolas-pacifier,OstovariRLNC} approaches. The first category aims at striking the best possible trade-off between energy expenditure (usually expressed by the number of transmissions) and delivery efficiency. On the other hand, the second category targets at $100\%$ packet delivery and treats the minimization of the related costs as a second priority task.

As we mentioned, in this work we focus on energy efficient broadcasting. The proposed approaches in this field can be further classified, based on the coding method, into: i) XOR-based, and ii) RLNC-based approaches. The first subclass of algorithms~\cite{CodeB,NCDS_conf,NCDS_journal,kunz-iwcmc,OstovariXOR,Directional} follows the concept of ``coding opportunity"~\cite{COPE} to combine multiple packets into an encoded one using bitwise XOR operations. The coding process is performed on a hop-by-hop basis, i.e. decoding takes place at every hop. The prominent algorithm of this subclass, CodeB~\cite{CodeB}, combines CDS-based broadcasting with XOR network coding. It also provides information exchange mechanisms that make possible the implementation in mobile environments. CodeB builds on top of the non coding PDP~\cite{PDP} scheme, which is actually a CDS-based broadcast algorithm. However, it can be directly applied to other CDS-based approaches. Wang et al. explore the benefits of employing XOR network coding on various underlying CDS-based broadcast schemes~\cite{NCDS_conf,NCDS_journal}. Moreover, the use of XOR coding over PDP and MPR~\cite{MPR}, two typical CDS based algorithms, in tactical networks has been studied in~\cite{kunz-iwcmc}.
In~\cite{OstovariXOR} the authors study the problem of broadcasting with deadlines in static ad hoc networks and propose alternative buffering schemes for the RAD mechanism. Finally, Yang and Wu~\cite{Directional} explore the benefits of XOR coding in energy efficient broadcasting when combined with directional antennas. 

The second subclass of the energy efficient category consists of RLNC-based algorithms~\cite{fragouli_rlnc,widmer-extreme-net,mahmood_arlnccf_icc,RLDP}. These algorithms build on the concepts of practical random linear network coding~\cite{PracticalNC}. Encoded packets are created using random linear combinations based on the theory of finite fields~\cite{linear,Ho_randomized}. Then, they are forwarded either probabilistically~\cite{fragouli_rlnc,widmer-extreme-net,mahmood_arlnccf_icc} or deterministically using a CDS-based algorithm~\cite{RLDP}. Coding is allowed only between packets in the same group, called ``packet generation'', and is performed in an end-to-end basis, i.e., encoded packets are decoded at the prospective receivers and only when a sufficient amount of encoded packet is gathered. These algorithms require special mechanisms for allowing nodes to distributively agree on the grouping of packets into generations.

In the second major category of network coding enabled broadcast algorithms the focus is on guaranteeing $100\%$ reliability. Here, there are also two subclasses that can be identified. The first consists mainly of XOR-based algorithms that adopt a rateless approach~\cite{DiSC,WangXOR,rahnavard2008CRBCast,vellambi2010FTS}. Algorithms that utilize rateless coding keep producing encoded packets until all receivers are capable of decoding the initial packets. As a result, these schemes require feedback information. In any case, implementing a feedback mechanism is not straightforward in mobile networks. Therefore, those algorithms are limited to static networks. The same holds for the second subclass of algorithms that also follow a rateless approach but utilize random linear network coding~\cite{widmer_rlnc-update,hou2008adapcode,cho_DRAGON,yang2011-RCODE-jrn,subramanian2012uflood,Chachulski-more,Koutsonikolas-pacifier,OstovariRLNC}. Furthermore, most algorithms in this subclass use only intra-source coding, i.e. encoding packets only from the same source which significantly limits the coding gains in a multi-source scenario~\cite{mahmood_arlnccf_icc}.

\section{Conclusion}\label{sec:Conclusion}
XOR-based coding has been successfully used to enhance the energy efficiency of broadcasting in mobile ad hoc networks. We demonstrated, through extensive experimentation, that using the M/U termination criterion in the baseline broadcast algorithm severely impairs performance in several cases. Unfortunately, we found that alternative termination criteria that are proposed in the literature are not capable of efficiently supporting the coding process. As a result, we introduced a novel termination criterion that is fully compatible with XOR-based network coding. Furthermore, we revised some of the coding internals to enhance performance and at the same time reduce complexity. More specifically, we proposed a lightweight method for detecting coding opportunities that operates without a reception table. We introduced the concept of ``Coded Redundancy'' that reduces the end-to-end delay by increasing the packet redundancy across the network at no additional cost. Finally, we improved the forwarder election process of the proposed algorithm by exploiting information that was originally used only for coding purposes. The efficiency of NOB-CR, the algorithm that incarnates all the aforementioned modifications, was demonstrated through extensive simulations.

\section*{References}
\bibliographystyle{elsarticle-num}
\bibliography{ref}

\end{document}